\documentclass[draft,11pt,reqno]{amsart}
\usepackage{amsmath,amsgen,amsbsy,amstext,amsopn,amsthm,amssymb}

\newtheorem{thm}{THEOREM}

\newtheorem{lem}{LEMMA}[section]

\theoremstyle{definition}
\newtheorem{rem}{REMARK}

\newcommand{\infspec}{{\rm inf\ spec\ }}
\newcommand{\R}{{\mathbb R}}

\newcommand{\C}{{\mathbb C}}
\newcommand{\ap}{\alpha\pi^{-1}}

\newcommand{\Da}{{D}^\ast}
\newcommand{\Ea}{{E}^\ast}
\newcommand{\Fa}{{F}^\ast}
\newcommand{\F}{{\mathcal F}}
\newcommand{\Em}{{\mathcal E}}
\newcommand{\Q}{{\mathcal Q}}
\newcommand{\Ll}{{\mathcal L}}

\renewcommand{\P}{{\mathcal P}}

\newcommand{\Hh}{{\mathcal H}}

\newcommand{\eps}{\varepsilon}

\newcommand{\aan}{a_{\lambda}}
\newcommand{\ac}{a^{\ast}_{\lambda}}
\newcommand{\E}{{\mathcal E}}
\newcommand{\ean}{\varepsilon^{\lambda}}
\newcommand{\half}{\mbox{$\frac{1}{2}$}}

\newcommand{\al}{{\alpha}}
\newcommand{\pa}{{\parallel}}

\newcommand{\so}{{\Sigma_\al}}
\newcommand{\vs}{ \sigma}

\newcommand{\as}{\sqrt{\alpha}}

\newcommand{\Ow}{{\mathcal O}}
\newcommand{\ora}{{|0\rangle}}
\newcommand{\lao}{{\langle 0|}}

\newcommand{\la}{\Lambda}

\newcommand{\mA}{{\mathcal A}}
\newcommand{\ua}{\uparrow}

\newcommand{\pe}{\psi_\eps}
\renewcommand{\l}{ \lambda}

\newcommand{\un}{{\mathchoice {\rm 1 \mskip-4mu l} {\rm 1\mskip-4mu
l} {\rm 1\mskip-4.5mu l} {\rm 1\mskip-5mu l}}}
\numberwithin{equation}{section}

\begin{document}

\title[Self-energy of one electron]
{Self-energy of one electron in  non-relativistic QED}
\author[I. Catto \& C. Hainzl]{Isabelle Catto and Christian Hainzl}
\address{CEREMADE, CNRS UMR 7534, Universit\'e Paris-Dauphine, Place
du Mar\'e\-chal de Lattre de Tassigny, F-75775 Paris Cedex 16,
France} \email{catto@ceremade.dauphine.fr}
\address{Mathematisches Institut, LMU M\"unchen,
Theresienstrasse 39, 80333 Munich, Germany}
\email{hainzl@mathematik.uni-muenchen.de}
\date{\today}

\date{\today}

\begin{abstract}
We investigate the self-energy of one electron  coupled to a
quantized radiation field by extending the ideas developed in
\cite{H}. We fix an arbitrary cut-off parameter $\la$ and recover
the $\al^2$-term of the self-energy, where
 $\alpha$ is the coupling parameter representing the fine structure
constant. Thereby we develop a method which allows to expand the
self-energy up to {\it any power} of $\al$. This implies that
perturbation theory is correct if $\la$ is fix.

As an immediate consequence we obtain enhanced binding for
electrons.
\end{abstract}

\maketitle

\section{INTRODUCTION AND MAIN RESULTS}

In recent times the self-energy of an electron was studied in
several articles. In \cite{LL}, Lieb and Loss showed that in the
limit of large cut-off parameter $\la$, perturbation theory is
conceptually wrong.

A different method of investigating   the self-energy was
developed in \cite{H}. Therein the cut-off parameter $\la$ was
fixed and the self-energy in the case of small coupling parameter
$\al$ was studied. It turned out that one photon is enough to
recover the first order in $\al$ which implies at the same time
that perturbation theory, in $\al$, is {\it correct} if $\la$ is
kept fix.

By similar methods Hainzl and Seiringer evaluated in \cite{HS} the
mass renormalization via the dispersion relation and proved that
after renormalizing the mass the binding energy of an electron in
the field of a nucleus, to leading order in $\al$, has a finite
limit as $\la$ goes to infinity.

As our main result in the present paper we recover the next to
leading order, the $\al^2$-term, of the self-energy of an
electron.

As a byproduct of the proof we develop a method which allows  to
expand the self-energy, step by step, up to {\it any power} of
$\al$.

As an immediate consequence of our main result we obtain enhanced
binding for electrons. This means that a dressed electron in the
field of an external potential $V$ can have a bound state even if
the corresponding Schr\"odinger operator $p^2 +V$ has only
essential spectrum. Enhanced binding for charged particles without
spin was previously proven in \cite{HVV}.

\subsection{Self-energy}

The self-energy of an electron is described as the bottom of the
spectrum of the so-called Pauli-Fierz operator
\begin{equation}\label{rpf}
T = (p + \sqrt{\al}A(x))^2 + \as\vs\cdot B(x) + H_f\;.
\end{equation}
acting on  the Hilbert space
\begin{equation*}
\Hh = \Ll^{2}({\mathbb {R}}^{3};\C^2)\otimes \F
\end{equation*}
where $\F=\bigotimes_{n=0}^{+\infty}\Ll_b^2(\R^{3n};\C^{2})$ is
the Fock space for the photon field  and $\Ll_b^2(\R^{3\,n})$ is
the space of symmetric functions in $\Ll^2(\R^{3\,n})$
representing $n$-photons states.

We fix units such that  $\hbar=c=1$ and the electron mass
$m=\half$. The electron charge is then given by $e=\as$, with
$\alpha\approx 1/137$ the fine structure constant. In the present
paper $\al$ plays the role of a small, dimensionless number which
measures the coupling to the radiation field. Our results hold for
{\it sufficiently small values } of $\al$. $\vs$ is the vector of
Pauli matrices $(\sigma_1,\sigma_2,\sigma_3)$. Recall that the
$\sigma_i$'s are hermitian $2\times 2$ complex matrices and
fulfill the anti-commutation relations $\sigma_i\sigma_j +
\sigma_j\sigma_i = 2\un_{\scriptscriptstyle \C^2} \delta_{i,j}$.
The operator $p= -i\nabla$ is the electron momentum while $A$ is
the magnetic vector potential. The magnetic field is $B = {\rm
curl} \ A$.

The vector potential is
\begin{equation*}
A(x) = \sum_{\lambda = 1,2} \int_{\R^3}
\frac{\chi(|k|)}{2\pi\,|k|^{1/2}} \,\ean(k)\big[ \aan(k) e^{ikx}
+\ac(k) e^{-ikx}\big] dk\;,
\end{equation*}
and  the corresponding magnetic field reads
\begin{eqnarray}\nonumber
B(x) &=&\sum_{\lambda = 1,2} \int_{\R^3}
\frac{\chi(|k|)}{2\pi\,|k|^{1/2}}\, (k\times i\ean(k))\big[
\aan(k) e^{ikx} -
  \ac(k) e^{-ikx}] dk\;,
\end{eqnarray}
where the annihilation and creation operators $\aan$ and  $\ac$,
respectively,  satisfy the usual commutation relations
\begin{equation*}
[a_\nu (k),\ac(q)] = \delta(k-q)\delta_{\lambda,\nu}\;,
\end{equation*}
and
\begin{equation*}
[\aan(k), a_\nu(q)] = 0, \ \ [\ac(k),a_\nu^\ast(q)] = 0\;.
\end{equation*}
The vectors $\ean(k) \in \R^3$ are orthonormal polarization
vectors perpendicular to $k$, and they are chosen in a such a way
that
\begin{equation}\label{convention}
    \varepsilon^2(k)=\frac{k}{| k|}\wedge
    \varepsilon^1(k)\;.
\end{equation}
The function $\chi(|k|)$ describes the ultraviolet cutoff on the
wave-numbers $k$. We choose for $\chi$  the Heaviside function
$\Theta(\Lambda -|k|)$. (More general cut-off functions would work
but let us nevertheless emphasize the fact that we shall sometimes
use the radial symmetry of  $\chi$ in the proofs.) Throughout the
paper we assume $\la$ to be an arbitrary but fixed positive
number.

The photon field energy $H_f$ is given by
\begin{equation}
H_f = \sum_{\lambda= 1,2} \int_{\R^3} |k| \ac (k) \aan (k) dk
\end{equation}
and the field momentum reads
\begin{equation}
P_f = \sum_{\lambda= 1,2} \int_{\R^3} k \ac (k) \aan (k) dk\;.
\end{equation}

In the following we use the notation
\begin{equation}
A(x)= D(x) + D^*(x), \, \, B(x)=E(x) + E^*(x)
\end{equation}
for the vector potential, respectively the magnetic field.

The operators $D^*$ and $E^*$ create a photon  wave function
$G(k)e^{-i k\cdot x} $ and $H(k)e^{-ik\cdot x}$, respectively,
where  $G(k)=(G^1(k),G^2(k))$ and $H(k)=(H^1(k),H^2(k))$ are
vectors of one-photon states, given by
\begin{equation}\label{defG}
G^{\lambda}(k)=\frac {\chi(|k|)}{2\pi |k|^{1/2}}\;\ean(k)\;,
\end{equation}
and
\begin{equation}\label{defH}
H^{\lambda}(k)=\frac {-i\chi(|k|)}{2\pi |k|^{1/2}} \;k\wedge
\ean(k) =-i\,k\,\wedge G^{\lambda}(k)\;.
\end{equation}

It turns out to be convenient to denote a general vector  $\Psi
\in \Hh$ as a direct sum
\begin{equation}
\Psi= \sum_{n\geq 0}\psi_n\;,
\end{equation}
where $\psi_n = \psi_n(x,k_1,\dots,k_n)$ is a $n$-photons state.
For simplicity, we do not include the variables corresponding to
the polarization of the photons and the spin of the electron.

{}From \cite{H} we know that the first order term in $\al$ of the
self-energy
\begin{equation}
\so = \infspec T
\end{equation}
is given by
\begin{equation}\label{isa}
 \ap \la^2 - \al\langle 0 | E\mA^{-1}
\Ea| 0\rangle=2\ap [\la - \ln(1+\la)]\;,
\end{equation}
where $\mA = P_f{}^2 + H_f$ and $|0\rangle $ is the vacuum in the
Fock space $\F$. Recall that the vacuum polarization, $\al \langle
0 | A^2 | 0\rangle = \ap \la^2$, enters somehow \textit{ab initio}
the game, whereas the second term in the r.h.s. of (\ref{isa})
stems from the magnetic field $B$. But now, for the next to
leading order $\al^2$ all terms contribute.

\begin{thm}[\textbf{Expansion of the self-energy up to
second \- order}]\label{thm1} Let $\la$ be fixed. Then, for $\al$
small enough,
\begin{multline}\label{diego}
\so =\al \Big[\pi^{-1}\Lambda^2-\langle 0 | E\mA^{-1} \Ea|
0\rangle\Big] - \al^2 \Big[\langle 0|D D \mA^{-1} \Da\Da |
0\rangle +
\\+\langle 0|E \mA^{-1}E \mA^{-1} \Ea \mA^{-1} \Ea| 0\rangle +
4\,\langle 0|E \mA^{-1}P_{f}\cdot D \mA^{-1} P_{f}\cdot\Da
\mA^{-1} \Ea| 0\rangle - \\-2\langle 0|E\mA^{-1}E\mA^{-1}
D^*D^*\ora- \langle 0 | E \mA^{-1} \Ea | 0\rangle \pa \mA^{-1} \Ea
| 0 \rangle\pa^2 \Big] +\\+ \Ow \big(\al^{5/2}\ln(1/\al)\big).
\end{multline}
\end{thm}
\begin{rem} Throughout the paper the notation $\Ow(f(\al))$ means that
there is a positive constant $C$ such that $|\Ow(f(\al))|\leq C\,
f(\al)$.
    \end{rem}
\subsection{Enhanced binding}

As an immediate consequence of Theorem \ref{thm1} we are able to
prove enhanced binding for electrons, which was already shown in
\cite{HVV} for charged bosons. Namely, if we take a negative
radial potential $V = V(|x|)$ with compact support such that $p^2
+ V$ has purely continuous spectrum, thus no bound-state, but a
so-called zero-resonance which satisfies the equation
\begin{equation}\label{zerorr}
  \psi(x) = - \frac 1{4\pi} \int\frac{ V(y) \psi(y)}{|x-y|} dy\;.
\end{equation}
Then after turning on the radiation field, even for infinitely
small coupling $\al$, the Hamiltonian
\begin{equation}
H_\al = T + V
\end{equation}
has a ground state. To this end we  use a result of \cite{GLL}
stating that the inequality
\begin{equation}
\infspec H_\al < \so
\end{equation}
guarantees the existence of a ground state. Earlier the existence
of a ground state, for small coupling, has been proven in
\cite{BFS}.
\begin{thm}[\textbf{Enhanced binding}]
    \label{cor1}
Let $V$ be a negative continuous function, which is radially
symmetric and with compact support. Assume that the corresponding
Schr\"{o}dinger operator $p^2 +V$ has no eigenvalue, but that
there exists a non-trivial radial solution of (\ref{zerorr}). Then
at least for small  values of $\al$  the operator $H_\al$ has a
ground state.
\end{thm}
Notice, due to the spin the ground state is twice degenerate
(\cite{HiSp1}). Earlier, in the dipole approximation enhanced
binding in the limit of large coupling $\al$  was shown in
\cite{HiSp2}.

\section{PROOF OF THEOREM \ref{thm1}}

We will follow the methods developed in \cite{H} and extend the
ideas therein. For sake of a simplified notation we introduce the
unitary transform
\begin{equation}
U = e^{iP_f \cdot x}
\end{equation}
acting on $\Hh$. Notice that $U\psi(x) = e^{ik\cdot x}\psi(x)$,
\[
U\big(E^*(x) \psi(x)\big) = H(k)\psi(x)\,
\]
and
\[U
(D^*(x)\psi(x)) = G(k)\psi(x)\;.
\]
More generally, for a $n$-photons component,  we have
\begin{multline*}
U\big(E^*(x) \psi_{n}(x,k_{1},\dots,k_{n})\big) =\\
=\frac{1}{\sqrt{n+1}}\sum_{i=1}^{n+1} H(k_{i})
\psi_{n}(x,k_{1},\dots,\check{k}_{i},\dots,k_{n+1})\,
\end{multline*}
and
\begin{multline*}
U\big(D^*(x) \psi_{n}(x,k_{1},\dots,k_{n})\big) =\\
=\frac{1}{\sqrt{n+1}}\sum_{i=1}^{n+1} G(k_{i})
\psi_{n}(x,k_{1},\dots,\check{k}_{i},\dots,k_{n+1})\,
\end{multline*}
where the notation $\check{\cdot}$ means that the corresponding
variable has been omitted. Since
\begin{equation}
UpU^*=p-P_f
\end{equation}
we obtain
\begin{equation}
UTU^* = \big(p-P_f + \as A\big)^2 + \as \vs \cdot B + H_f\;,
\end{equation}
where $A=A(0)$ and $B=B(0)$.

Obviously,
\begin{equation}
\infspec\big[UTU^*] = \infspec T\;.
\end{equation}
Therefore in the following  we will rather work with $UTU^*$ which
we still denote by $T$.

We also introduce the notation
\begin{equation}
L= (p-P_f)^2 + H_f\;,
\end{equation}
\begin{equation}
\P = p-P_f\;,
\end{equation}
\begin{equation}
\Fa_f= 2\,\P_f\cdot \Da+ \vs\cdot \Ea\;
\end{equation}
and
\begin{equation}
\Fa= 2\,\P\cdot \Da+ \vs\cdot \Ea\;.
\end{equation}

Recalling that
\begin{equation}
A^2 = \la^2 \pi^{-1}+ 2 \Da D + \Da \Da + DD\;,
\end{equation}
we then have, for any general $\Psi \in \Hh$,
\begin{eqnarray}
(\Psi,T\Psi) &=&  \Lambda^2 \ap \pa \Psi \pa^2 + \|
p\,\psi_0\|^2+2\al \sum_{n\geq 1} (\psi_n,D^\ast D\psi_n) +\nonumber\\
&&+\Em_0[\psi_0,\psi_1] + \sum_{n \geq 0}
\Em[\psi_n,\psi_{n+1},\psi_{n+2}]\;,\label{Tpp}
\end{eqnarray}
where, as in \cite{H},
\begin{equation}\label{defE0}
\Em_0 [\psi_0,\psi_1] = ( \psi_1, L \psi_1)  + 2\,\as\,{\Re} (\Fa
\psi_0,\psi_1)
\end{equation}
and
\begin{eqnarray}\label{defE}
\Em[\psi_n,\psi_{n+1},\psi_{n+2}]&=&(\psi_{n+2} , L
\psi_{n+2})+\nonumber\\
&&+2\,\Re\left(\as\,\Fa \psi_{n+1}+ \al\Da\Da \psi_n,
\psi_{n+2}\right).
\end{eqnarray}

For simplicity, in this section, we shall actually work in the
momentum representation of the electron space. A $n$-photons
function $\psi_{n}$ will then be looked at as $\psi_n(l,k)$ with
$k = (k_1,...,k_n)$, where $l$ stands for the momentum variable of
the electron and is obtained from the position variable $x$ by
Fourier transform. In that case $\P$ is simply a multiplication
operator, and for short we use
\begin{equation}
\P \psi_n(l,k_1,\dots,k_n) = \Big(l-\sum_{i=1}^n k_i\Big) \psi_n
=: \P_n \psi_n\;,
\end{equation}
and similarly
\begin{equation}
H_f \psi_n(l,k_1,\dots,k_n) = \sum_{i=1}^n |k_i|\psi_n =: H_f^n
\psi_n\;.
\end{equation}

\subsection{Upper bound for $\so$}

As usual the trick is to exhibit a cleverly chosen trial function.
In \cite{H}, the leading order term in $\al$ is obtained by a
trial function $\overline{\Psi}^{(n)}$  with only one photon. The
idea to get the second order term is to add a 2-photons component
whose $\Ll^2$ norm is of the order of $\al$. More precisely, we
define the sequence of trial wave functions
\begin{eqnarray}
\Psi^{(n)}
&=&\overline{\Psi}^{(n)}+\al\,f_{n}\ua\otimes\mA^{-1}[\vs \cdot\Ea
+2P_{f}\cdot\Da]\mA^{-1}
\vs \cdot\Ea \ora -\nonumber\\
&&-\al\,f_{n}\ua\otimes\mA^{-1}\Da\Da\ora\;,\label{tf2}
\end{eqnarray}
with $\ua$ denoting the spin-up vector $(1,0)$ in $\C^{2}$, $f_n
\in H^{1}(\R^3;\R)$, $\pa f_n\pa=1$ and $\pa p f_n \pa \to 0$ when
$n$ goes to infinity, and where
\begin{equation}\label{tf}
\overline{\Psi}^{(n)}= f_n \ua\otimes| 0\rangle -\as f_n
\ua\otimes \mA^{-1}\vs\cdot\Ea\ora\;.
\end{equation}
Let us already observe that the choice for the trial function will
appear more natural after the proof of the lower bound (see below
the expected decomposition (\ref{defh1}) and (\ref{defhn})-with
$n=0$- of a two-photons state close to the ground state).

We are going to check that
\begin{equation}\label{zinedine}
     \lim_{{n\to+\infty}}\frac{(\Psi^{(n)},T\Psi^{(n)})}{\pa\Psi^{(n)}\pa^{2}}=
     \mathcal{E}_{1}\alpha+\mathcal{E}_{2}\alpha^{2}+\Ow(\al^{3})\;,
\end{equation}
where
\begin{equation}
     \label{E1}
     \mathcal{E}_{1}=\pi^{-1}\la^{2}-\lao E\mA^{-1}\Ea \ora\;,
\end{equation}
and
\begin{eqnarray}
\mathcal{E}_{2}&=&-\langle 0|D D \mA^{-1} \Da\Da | 0\rangle
-\langle 0|E \mA^{-1}E \mA^{-1} \Ea
\mA^{-1} \Ea| 0\rangle -\nonumber\\
&&- 4\,\langle 0|E \mA^{-1}P_{f}\cdot D
\mA^{-1} P_{f}\cdot\Da \mA^{-1} \Ea| 0\rangle- \nonumber\\
&& +2\langle 0|E\mA^{-1}E\mA^{-1} D^*D^*\ora+ \langle 0 | E
\mA^{-1} \Ea | 0\rangle \pa \mA^{-1} \Ea | 0 \rangle\pa^2
\end{eqnarray}
respectively denote the coefficient of $\al$ and $\al^{2}$ in
(\ref{diego}).

We first  point out that, for any $N$-photons wave function
$\varphi_{N}$, we have
\begin{equation}
     L(f_{n}\otimes \mA^{-1}
     \varphi_{N})-f_{n}\otimes\varphi_{N}\longrightarrow 0 \textrm{ in }
     H^{-1}(\R^3;\R)\otimes \Ll^{2}(\R^{3},\C^{2})^{N}-weak,\label{LA-1}
\end{equation}
as $n$ goes to infinity in virtue of the fact that
$\lim_{n\to+\infty}\| p f_n \|= 0$, and  since, by definition of
$L$ and $\mA$,
\begin{equation}
     L(f_{n}\otimes \mA^{-1}
     \varphi_{N})=f_{n}\otimes\varphi_{N}-2pf_{n}\otimes
P_{f}\mA^{-1}\varphi_{N}
     +p^{2}f_{n}\otimes \mA^{-1}\varphi_{N}\;.
\end{equation}
Then, with the help of (\ref{Tpp}) and the fact that $\|
f_{n}\|=1$, easy calculations yield
\begin{subequations}\label{limsup}
\begin{eqnarray}
\lefteqn{(\Psi^{(n)},T\Psi^{(n)})=}\nonumber\\
&=& \ap \la^{2}\pa \Psi ^{(n)}\pa^2+\pa p f_{n}\pa^{2}+ 2 \al \| D
\psi_{1}^{(n)}\|^{2}
+ 2 \al \| D \psi_{2}^{(n)}\|^{2}+\nonumber\\
&&+(\psi_{1}^{(n)},L \psi_{1}^{(n)})+2\as\,\Re (\Fa
f_{n}\ua,\psi_{1}^{(n)})+\label{limsup0}\\
&&+(\psi_{2}^{(n)},L \psi_{2}^{(n)})+2\as \,\Re
(\Fa\psi_{1}^{(n)}, \psi_{2}^{(n)})+2\al \Re (\Da\Da
f_{n}\ua,\psi_{2}^{(n)})=
\nonumber\\
&=&\ap \la^{2}\pa \Psi ^{(n)}\pa^2
-\al \langle 0| E\mA^{-1}\Ea\ora +o_{n}(1)+ \Ow(\al^{3}) +\nonumber\\
&&+2 \al^{2} \| D \mA^{-1}\sigma\ua\cdot \Ea\ora\|^{2}-\al^{2}\lao
D D \mA^{-1}\Da\Da \ora -
\label{limsup1}\\
&&-\al^{2}\lao \vs\ua\cdot E
\mA^{-1}F_f\mA^{-1}\Fa_f\mA^{-1}\vs\ua\cdot\Ea \ora+\label{limsup3}\\
&&+2\al^{2}\Re(L^{-1}\Fa\mA^{-1}\vs\ua\cdot\Ea f_{n},\Da\Da
f_{n}\ua)\;,\label{limsup4}
\end{eqnarray}
\end{subequations}
where $o_{n}(1)$ refers to a quantity that goes to $0$ as $n$ goes
to infinity and is some error term coming from the fact that
$\lim_{n\to+\infty}\| p f_n \|= 0$, while $\Ow(\al^{3})$ comes
from the $\al \| D\psi_{2}^{(n)}\|^{2}$ term. The proof of the
fact that
\begin{equation*}
     (\ref{limsup0})=-\al \langle 0| E\mA^{-1}\Ea\ora +o_{n}(1)
     \end{equation*}
is detailed in \cite{H}. We first check that $D\psi_{1}^{(n)}=0$,
or, equivalently,
\begin{equation*}
  D \mA^{-1}\sigma\ua\cdot \Ea\ora=0\;.
\end{equation*}
This simply follows  from the relation
\begin{equation}
\sum_{\lambda=1,2} \eps^{\lambda}_i  \eps^{\lambda}_j =
\delta_{i,j} - \frac{k_i\,k_j}{|k|^2}\;,
\end{equation}
and the obvious observation that, for every $i\in\{1,2,3\}$,
\[
D_{i}\mA^{-1}\sigma\ua\cdot
\Ea\ora=\sum_{j=1}^{3}\sigma_{j}\ua\sum_{\lambda=1,2}\int_{\R^{3}}
\frac{G_{i}^\lambda(k)\,H_{j}^\lambda(k)}{| k|^{2}+| k|}\,dk \;,\]
with the three vectors $\sigma_{j}\ua$, $j=1,2,3$, being linearly
independent. Then, if $\mathbf{{\epsilon^{j \, l\, n}}}$ denotes
the totally antisymmetric epsilon-tensor, we obtain, for every
$i,\ j\in\{1,2,3\}$,
\begin{multline}\label{epstens}
\sum_{\lambda =1,2} \int_{\R^{3}} \frac{G^{\lambda}_i (k)
H^\lambda_j (k)}{|k|^2 + |k|}dk = \sum_{\lambda =1,2}
\sum_{l,n=1}^{3}i\int_{\R^{3}} \frac{\chi(|k|)
\eps^\lambda_i(k)\big[ \mathbf{{\epsilon^{j \, l\, n}}}\,
\eps^\lambda_l(k) k_n\big]}{|k|^3 +
|k|^2}\,dk\\
=  \sum_{l,n=1}^{3}i\int_{\R^{3}} \frac{\chi(|k|)\big[\delta_{i,l}
- \frac{k_i\,k_l}{|k|^2}\big] \mathbf{{\epsilon^{j \, l\, n}}}
k_n}{|k|^3 + |k|^2}\,dk = 0\;.
\end{multline}

Concerning (\ref{limsup4}), we use the anti-commutation relations
of the $\vs_{j}$'s and the fact that  the functions
$H^{\lambda}(k)$ belong to $(i\R)^{3}$ while $G^{\lambda}(k)$
belong to $\R^{3}$ to check that
\begin{equation*}\Re(L^{-1}\P\cdot \Da\mA^{-1}\vs\cdot\ua\cdot\Ea f_n, \Da\Da f_n\ua)=o_n(1)
\;,
\end{equation*}and to deduce that
\begin{equation*}\label{limsup42}
(\ref{limsup4})= 2\al^{2}\pa f_n\pa^2\,\lao  E \mA^{-1}E
\mA^{-1}\Da\Da\ora+o_n(1) \;.
\end{equation*}

We now turn to (\ref{limsup3}) and  check  that
\begin{multline}\label{limsup32}
(\ref{limsup3})=-\al^{2}\langle 0|E \mA^{-1}E \mA^{-1} \Ea
\mA^{-1} \Ea| 0\rangle\\- 4\al^{2}\,\langle 0|E \mA^{-1}P_{f}\cdot
D \mA^{-1} P_{f}\cdot\Da \mA^{-1} \Ea| 0\rangle \;,
\end{multline}
since the cross term $ \Re \langle 0|E \mA^{-1}P_{f}\cdot D
\mA^{-1}\Ea \mA^{-1} \Ea| 0\rangle$ vanishes thanks again to the
fact that $G$ is real valued while $H$ is purely imaginary.

The last second-order term which appears in (\ref{diego}) is
easily recovered, once we have observed from \eqref{tf2} and
(\ref{tf}) that
\begin{equation*}
\|\Psi^{(n)}\|^{2}=1+\al\| \mA^{-1} \Ea\ora\|^2+\Ow(\al^2)\;.
\end{equation*}
Hence (\ref{zinedine}), by  dividing the l.h.s. of (\ref{limsup})
by $\|\Psi^{(n)}\|^{2}$.
\subsection{Lower bound for $\so$}
The proof will be divided into two steps. First, in
Subsection~\ref{ssec1}, we deduce  \textit{a priori } estimates
for any state which is ``close enough" to the ground state energy.
Next in Subsection~\ref{ssec2} we use these estimates to recover
the $\al^2$-term of the self-energy.

\subsubsection{A priori estimates}\label{ssec1}

Our first step will consist in improving a bit further the
estimates in~\cite{H}. Indeed, we may choose a state $\Psi$ in
$\Hh$, close enough to the ground state, such that $\pa\Psi\pa=1$
and
\begin{equation}
\so \leq (\Psi,T\Psi)\leq \so+C\al^{2} \leq \ap \la^2 - \al\langle
0 | E\mA^{-1} \Ea| 0\rangle + C\al^{2},
\end{equation}
where, here and below, $C$ denotes a positive constant that is
independent of $\al$ (but that might possibly dependent on $\la$).
We thus have as in  \cite{H}
\begin{equation}
     \label{bornesH}
\sum_{n\geq 0}(\psi_{n},L\, \psi_{n}) \leq C\,\al\;,
\end{equation}
hence
\begin{equation}
\sum_{n\geq 0}(\psi_{n},(\Da D+\Ea E) \psi_{n}) \leq C\,\al\;,
\end{equation}
in virtue of \cite[Lemma A.4]{GLL}. We now observe that
\begin{equation}\label{E0}
    \Em_0 [\psi_0,\psi_1] =-\al \| L^{-1/2}\Fa\psi_{0} \|^{2} +(
h_1, L h_1)\;,
\end{equation}
where
\begin{equation}\label{defh1}
\psi_{1}=-\as\,L^{-1} \Fa\psi_{0}+h_{1}\;,
\end{equation}
and that, for every $n\geq 0$,
\begin{eqnarray}\label{E}
\Em[\psi_n,\psi_{n+1},\psi_{n+2}]&= & -\al \|
L^{-1/2}\Fa\psi_{n+1} +\as
L^{-1/2}\Da\Da\psi_{n}\|^{2}\nonumber\\
&&+( h_{n+2}, L h_{n+2})\;,
\end{eqnarray}
where
\begin{equation}\label{defhn}
\psi_{n+2}=-\as\,L^{-1}
\Fa\psi_{n+1}-\al\,L^{-1}\Da\Da\psi_{n}+h_{n+2}\;.
\end{equation}
Comparing with (\ref{Tpp}), we thus rewrite
\begin{subequations}\label{min1}
\begin{eqnarray}
(\Psi,T\Psi)
  &=&\al \Lambda^2 \pi^{-1} \pa \Psi \pa^2-\al \|
L^{-1/2}\Fa\psi_{0} \|^{2}-\label{min11}\\
  && -\al \sum_{n\geq 0}\| L^{-1/2}\Fa\psi_{n+1} +\as
L^{-1/2}\Da\Da\psi_{n}\|^{2}+\label{min12}\\
  &&+ \| p\,\psi_0\|^2+2\al \sum_{n \geq 1} (\psi_n,D^\ast D\psi_n)
+\sum_{n\geq 1} (h_{n},L\,h_{n}).\label{min13}
\end{eqnarray}
\end{subequations}

Our first step will consists in observing that the estimates in
\cite{H} yield
\begin{equation}
     \sum_{n\geq 1} (h_{n},L\,h_{n})\leq C\,\al^{2}
     \label{borneh}
\end{equation}
and
\begin{equation}
     \pa p\,\psi_{0}\pa^{2}\leq C\,\al^{2}\;,
     \label{bornep0}
\end{equation}
thereby  improving the estimate on the zeroth order term in
(\ref{bornesH}). These bounds will follow from the fact that only
the terms in the first two lines of (\ref{min1}) contribute to
recover the first to leading order term up to $\Ow(\al^2)$. Hence,
all the (positive) terms in (\ref{min13}) are at most of the order
of $\al^2$. \vskip6pt Indeed, on the one hand, we recall from [H]
that
\begin{equation*}
     \left| \al (\vs\cdot \Ea \psi_{0}, L^{-1}\vs\cdot \Ea\psi_{0} )
     -\al \pa \psi_{0}\pa^{2}\,
     \langle 0 | E\mA^{-1}
\Ea| 0\rangle\right| \leq C\,\al\, \pa p \psi_{0}\pa^{2}\;,
\end{equation*}
\begin{equation*}
     \Re (\vs\cdot\Ea \psi_{0}, L^{-1} \P\cdot \Da\psi_{0})=0\;,
\end{equation*}
and
\begin{equation*}
      \al(\P\cdot \Da\psi_{0}, L^{-1} \P\cdot \Da\psi_{0})
      \leq C\,\al\,\pa p \,\psi_{0}\pa^{2}\;.
\end{equation*}
Hence
\begin{equation} \label{contriE0}
\left| \al \| L^{-1/2}\Fa\psi_{0}\|^{2}
     -\al \pa \psi_{0}\pa^{2}\,
     \langle 0 | E\mA^{-1}
\Ea| 0\rangle\right| \leq C\,\al\, \pa p \psi_{0}\pa^{2}\;.
\end{equation}
Therefore, concerning the last term in \eqref{min11}, we have
\begin{equation}
\label{term1} -\al \| L^{-1/2}\Fa\psi_{0}\|^{2}
     =-\al \pa \psi_{0}\pa^{2}\,
     \langle 0 | E\mA^{-1}
\Ea| 0\rangle+\Ow(\al^2)\;,
\end{equation}
thanks to (\ref{bornesH}). \vskip6pt On the other hand, we now
estimate the different terms in (\ref{min12}), for every $n\geq
0$. More precisely,
\begin{subequations}
\begin{eqnarray}
\eqref{min12}&=&-\al \| L^{-1/2}\Fa\psi_{n+1} +\as
  L^{-1/2}\Da\Da\psi_{n}\|^{2}=\nonumber\\
  &=&- \al \| L^{-1/2}\Fa\psi_{n+1}\|^{2}-
  \al^2  \,(\psi_{n},D\,D L^{-1}\Da\Da\psi_{n})-\label{diag3}\\
&&- 2\al^{3/2}\, \Re (\Fa \psi_{n+1}, L^{-1}\Da
\Da\psi_{n})\;.\label{cross2}
\end{eqnarray}
\end{subequations}
It is shown in \cite{H}, that
\begin{equation}\label{ordre1}
    \Big|\| L^{-1/2}\Fa\psi_{n+1}\|^{2}-
    \pa\psi_{n+1}\pa^{2}\lao E\mA^{-1}\Ea\ora\Big|\leq C\,(\psi_{n+1},
    L\psi_{n+1})\;.
 \end{equation}
This follows from the three bounds
\begin{multline}\label{equh1}
  \left|
     (\vs\cdot \Ea \psi_{n+1}, L^{-1}\vs\cdot \Ea\psi_{n+1})
     - \pa\psi_{n+1}\pa^{2}\langle 0 | E\mA^{-1}
\Ea| 0\rangle\right|\\
\leq C\,(\psi_{n+1}, L\psi_{n+1})\;,
\end{multline}
\begin{equation*}
     (\P\cdot \Da \psi_{n+1}, L^{-1}\P\cdot \Da\psi_{n+1})
     \leq C\,(\psi_{n+1}, L\psi_{n+1})\;,
\end{equation*}
and
\begin{equation*}
      | \Re(\P\cdot \Da \psi_{n+1}, L^{-1}\vs\cdot \Ea\psi_{n+1})|
     \leq C\,(\psi_{n+1}, H_{f}\psi_{n+1})\;,
\end{equation*}
whose proofs are  detailed  in \cite{H}. (See also the proof of
Lemma~\ref{lem:B1} in Appendix B below, which follows the same
patterns.) Moreover, from Lemma~2 in the Appendix of \cite{H},
\begin{multline}
\label{evalE3}
  \left|
     \al^{2} (\psi_{n},D\,D L^{-1}\Da\Da\psi_{n})
     -\al^{2} \pa\psi_{n}\pa^{2}\,\langle 0 | D\, D \mA^{-1}
\Da\Da| 0\rangle\right|\\
\leq C\,\al^{2}(\psi_{n+1}, L\psi_{n+1})\;.
\end{multline}
Actually, only the upper bounds of (\ref{equh1}) and
(\ref{evalE3}) are proven in \cite{H} which indeed  suffices for
the first order term, but following the methods described in
Appendix B the estimates (\ref{equh1}) and (\ref{evalE3}) are
easily derived.

For (\ref{cross2}), we get from the proof of Lemma~\ref{lem:C2} in
Appendix~C below
\begin{multline}
\label{cross22}
     \al^{3/2} | (\Fa \psi_{n+1}, L^{-1}\Da\Da\psi_{n})|
     \leq \\
     \leq C\,\al^{2}\,\pa \psi_{n}\pa^{2}+C\,\alpha (\psi_{n+1}, L\psi_{n+1})+C\,\al\,(\psi_{n}, L\psi_{n})\;.
\end{multline}
Summing up \eqref{ordre1}, \eqref{evalE3} and \eqref{cross22} over
$n\geq 0$ and using \eqref{term1} and  (\ref{bornesH}), we first
deduce from \eqref{min1} that
\begin{eqnarray*}
\lefteqn{\al\pi^{-1}\la^{2}-\al \pa \Psi\pa^{2}\,
     \langle 0 | E\mA^{-1}
\Ea| 0\rangle +\Ow(\al^2)\geq}\hskip2truecm\nonumber\\
&\geq&\so
\geq (\Psi,T\Psi)+\Ow(\al^2)=\\
&=&\ap \Lambda^2  \pa \Psi \pa^2-\al \pa \Psi\pa^{2}\,
     \langle 0 | E\mA^{-1}
\Ea| 0\rangle +\Ow(\al^2)+\nonumber\\
&&+ \|
  p\,\psi_0\|^2+2\al \sum_{n \geq  1}(\psi_n,D^\ast D\psi_n)
+\sum_{n\geq 1} (h_{n},L\,h_{n})\;.
\end{eqnarray*}
Whence (\ref{borneh}) and (\ref{bornep0}). \vskip10pt We now make
use of these bounds to derive the second order terms in
(\ref{diego}).

\subsubsection{Recovering the $\al^2$-terms.} \label{ssec2}
As a first consequence of (\ref{bornep0}), we deduce from
(\ref{contriE0}) that
\begin{equation}
\label{step1} -\al \| L^{-1/2}\Fa\psi_{0}\|^{2}
     =-\al \pa \psi_{0}\pa^{2}\,
     \langle 0 | E\mA^{-1}
\Ea| 0\rangle+\Ow(\al^{3})\;.
\end{equation}
\vskip10pt It turns out that, although it was not necessary
hitherto, we now have to introduce an infrared regularization as
in~\cite{HS} to deal with the terms in (\ref{min12}) (or
equivalently in \eqref{diag3} and \eqref{cross2}). Therefore, in
the definition (\ref{defE}) of $\E$ we replace the operator $L$ by
\[
L_{\alpha}\equiv L+\alpha^{3}\;,
\]
and the extra term $\al^{3} \sum_{n\geq 2}\pa \psi_{n}\pa^{2}$
contributes as an additional $\Ow(\al^{3})$ in (\ref{Tpp}). The
definition of $h_{n+1}$ has of course to be modified accordingly
by replacing $L^{-1}$ by $L_{\al}^{-1}$ in (\ref{defhn}). We shall
nevertheless keep the same notation for $h_{n+1}$, and we also
emphasize the fact that the bound (\ref{borneh}) obviously remains
true. \vskip6pt Keeping this minor modification in mind, we  now
go back to (\ref{min1}) and we shall now use the decompositions
(\ref{defh1}) and (\ref{defhn}) of $\psi_{n+1}$, $n\geq 1$, in
terms of $\psi_{n}$, $\psi_{n-1}$ and $h_{n+1}$ to exhibit the
remaining second order terms, as guessed from the upper bound.
\vskip6pt More precisely, the following quantity is now to be
estimated
\begin{subequations}
\label{xian}
\begin{eqnarray}
\lefteqn{ -\al \| L_\al^{-1/2}\Fa\psi_{n+1}+\as
  L_\al^{-1/2}\Da\Da\psi_{n}\|^{2}=}\nonumber\\
&=&-\al \| L_\al^{-1/2}\Fa h_{n+1}\|^2 -\al^2\| L_\al^{-1/2}\Fa
L_\al^{-1}\Fa\psi_{n}\|^2 -\label{xian1}\\
  && -\al^2\| L_\al^{-1/2}\Da \Da\psi_{n}\|^2 -\al^3\| L_\al^{-1/2}\Fa
L_\al^{-1}\Da\Da\psi_{n-1}\|^{2}+\label{xian2}\\
&&+2\al^2 \Re(L_\al^{-1}\Fa L_\al^{-1}\Fa\psi_{n},\Da\Da\psi_n)+\label{xian3}\\
&&+2\al^{3/2} \Re (L_\al^{-1}\Fa L_\al^{-1}\Fa\psi_{n},\Fa h_{n+1})-\label{xian4}\\
&& - 2\al^{3/2} \Re (L_\al^{-1}\Da\Da\psi_{n},\Fa h_{n+1})-
  \label{xian5}\\
  &&-2\al^{5/2} \Re (L_\al^{-1}\Fa L_\al^{-1}\Fa\psi_{n},\Fa L_\al^{-1}\Da\Da\psi_{n-1})
  -\label{xian6}\\
   &&-2\al^{5/2} \Re (L_\al^{-1}\Fa L_\al^{-1}\Da\Da\psi_{n-1},\Da\Da\psi_{n})
  +\label{xian7}\\
  &&+2\al^{2} \Re (L_\al^{-1}\Fa L_\al^{-1}\Da\Da\psi_{n-1},\Fa
  h_{n+1})\label{xian8}\;,
\end{eqnarray}
\end{subequations}
with here and below the convention that the terms containing
$\psi_{n-1}$ vanish for $n=0$. \vskip6pt

In order to lighten the presentation, the sequel of the proof has
been organized as follows. The contributing terms in (\ref{xian1})
and (\ref{xian3}) are investigated in Appendix B and the terms in
(\ref{xian4})--(\ref{xian8}) are shown to be of higher order in
Appendix C.

Admitting these lemmas for a while, we thus have from
Lemma~\ref{lem:xian1} and  Lemma~\ref{lem:B2} in Appendix~B below
and \eqref{bornesH} and \eqref{borneh},
\begin{multline}\label{terme1}
\eqref{xian1}=-\al \bigl(1-\| \psi_{0}\|^{2}
\bigr)\,\lao E\mA^{-1}\Ea\ora +\\
+\al^{2}\lao E\mA^{-1} \Ea\ora \;\| \mA^{-1}\Ea\ora \|^{2}-\al^2\,
\lao E \mA^{-1}E\mA^{-1}\Ea \mA^{-1}\Ea\ora -\\
 -4\al^2\,\lao E \mA^{-1}\P_{f}\cdot
D\mA^{-1}\P_{f}\cdot\Da
\mA^{-1}\Ea\ora+\Ow\big(\al^{5/2}\ln(1/\al)\big)\;.
    \end{multline}

{}From (\ref{evalE3}) and (\ref{bornesH}) again, we identify the
second order term in \eqref{xian2}; namely,
\begin{equation} \label{diag22}
\eqref{xian2}
    =-\al^{2} \,\langle 0 | D\, D \mA^{-1}
\Da\Da| 0\rangle+\Ow(\al^3)\;,
\end{equation}
since the second term in \eqref{xian2} is easily checked to be
$\Ow(\al^3)$. (Note that \eqref{evalE3} remains true when $L$ is
replaced by $L_\al$.)

The last contributing terms follows from Lemma~\ref{lem:B3} and
\eqref{bornesH}
\begin{equation}
\label{terme2} \eqref{xian3}=2\al^2\,\lao E \mA^{-1}E\mA^{-1}\Da
\Da\ora+\Ow(\al^{5/2}\ln(1/\al))\;.
\end{equation}
Finally, using the \textit{a priori} estimates \eqref{bornesH} and
\eqref{borneh}, and with the help of Lemma~\ref{lem:C1} to
Lemma~\ref{lem:C5}, we deduce that
\begin{multline}\label{errors}
\eqref{xian4}+\eqref{xian5}+\eqref{xian6}+\eqref{xian7}+\eqref{xian8}=\Ow\big(\al^{5/2}\ln(1/\al)\big)\;.
\end{multline}
 To deduce \eqref{diego}  we go back to \eqref{min1}. We simply bound from
below the terms in \eqref{min13} by zero, and identify
\eqref{min11} and \eqref{min12},  by using  \eqref{step1} and by
inserting  \eqref{terme1}, \eqref{diag22}, \eqref{terme2} and
\eqref{errors} in \eqref{xian}.
\begin{rem}
It would be possible to improve the error estimates to
$\Ow(\al^3)$, but we do not want to overburden the paper with too
many estimates. We just mention as an example that, from the proof
of the upper bound, we know that we may choose a state $\Psi$ in
$\Hh$, close enough to the ground state, such that $\pa\Psi\pa=1$
and
\begin{equation*}
\so \leq (\Psi,T\Psi)\leq \so+C\al^{3} \leq \ap \la^2 +
\al\E_1+\al^2\E_2+\Ow(\al^3)\;.
\end{equation*}
Then, arguing as in Subsection \ref{ssec1}, we infer from
\eqref{min1} that actually
\begin{equation}\label{borneh1}
\sum_{n\geq 0}\big(h_{n+1},Lh_{n+1}\big)+\|p\,\psi_{0}\|^2\leq
C\,\al^{5/2}\ln(1/\al)\;.
\end{equation}
This new and better bound now helps to improve all error estimates
on quantities  which involve $h_{n+1}$ and $\pa p\psi_0\pa^2$
(like \eqref{contriE0}, for example), and so on by a kind of
bootstrap argument.
\end{rem}
\begin{rem}
By means of the methods developed throughout the proof it is now
possible to expand the self-energy up to   any power of $\al$, but
unfortunately the number of estimates rapidly increase. We know from
perturbation theory that to gain the  $\al^3$-term we just need to add the 
term \begin{equation}\label{aldrei}
-\as \mA^{-1} (F +F^*)\psi_2 - \al \mA^{-1} D^*D^*\psi_1
\end{equation}
and  normalize the corresponding state. The 1- and 2-photon parts $\psi_1 $ and $\psi_2$ are defined in the upper bound (see (\ref{tf2})).
Notice that (\ref{aldrei}) also includes the 1-photon term
$\al^{3/2} \mA^{-1} F\big( \mA^{-1} F^* \mA^{-1}E^*  +
\mA^{-1}D^*D^* \big)\ora$.

\end{rem}

\vskip10pt

\section{PROOF OF THEOREM \ref{cor1}}
To prove the Theorem we will proceed similarly to \cite{HVV} and
check the binding condition of \cite{GLL} for $ H_\al$. Namely, we
will show that
\begin{equation} \label{bindcon}
\infspec H_\al < \so - \delta \al^2 +
\Ow\big(\al^{5/2}\ln(1/\al)\big),
\end{equation}
for some positive constant $\delta$. To this end we define a one
and a two-photons state similar to the previous section to recover
the self-energy, and  we  add an extra appropriately chosen
one-photon component which involves the gradient of  an electron
function which is close to a zero-resonance state; that is,  a
radial solution of the equation
\begin{equation}\label{zeror}
\psi(x) = -\frac{ 1}{4\pi} \int \frac{V(y) \psi(y)}{|x-y|} dy.
\end{equation}
Let $r_0$ denote the radius of the support of $V$, then, due to
Newton's theorem,
\begin{equation}
\psi(x) = \frac C{|x|}
\end{equation}
for $|x| \geq r_0$ and an appropriate constant $C$. Notice that
$\psi$ satisfies
\begin{equation}
-\Delta \psi + V(x)\psi = 0\;.
\end{equation}
Due to elliptic regularity properties (see e.g. \cite{LL2}), we
infer that $\psi \in C^2(\R^3)$. \vskip6pt To make  $\psi$ an
$\Ll^2$-function we are going to truncate it. It turns out to be
reasonable to do so at distance $|x| \sim 1/\al$ from the origin.
To this end we take functions $u(t),v(t)$ $\in C^2(\R)$ with $u^2
+ v^2 =1$ and $u =1$ for $t \in [0,1] $ and $u=0$ for $t \geq 2$,
and we define
\begin{equation}
\psi_\eps(x) = \psi(x) u(\eps \al |x|).
\end{equation}
Assume $1/(\eps \al) \geq 2r_0$, so
\begin{equation}
  \psi_\eps (x)=\frac {C}{|x|}u(\eps \al |x|)
\end{equation}
for $|x| \geq r_0$. Therefore we may find positive  constants
$C_{1}$ and $C_{2}$, depending on  $r_0$, such that
\begin{equation}\label{pein}
\pa p^2 \psi_\eps\pa^2 \leq C_1 \pa p \psi_\eps\pa^2 \leq \al \eps
C_2 \pa \psi_\eps\pa^2.
\end{equation}
Notice that  $\|\psi_\eps\|^2 = C\, (\al\eps)^{-1}$.

Throughout the previous section we have worked with the operator
$A(0)$. Here, the Hamiltonian also depends on the electron
variable $x$. In order to adapt the method developed in the
previous section we introduce again the unitary transform
\begin{equation}
U= e^{i P_f\cdot x}
\end{equation}
acting on the Hilbert space $\Hh$. When applied to a $n$-photons
function $\varphi_n$ we obtain $U\varphi_n= e^{i(\sum_{i=1}^n
k_i)\cdot x} \varphi_n(x,k_1,\dots,k_n)$.

Since $U p U^* = p - P_f$ we infer the corresponding transform for
the Hamiltonian $H_\al$
\begin{equation}
U H_\al U^* = (p - P_f +\as A)^2 +   \as \vs\cdot B+ H_f +V(x)\;,
\end{equation}
which we denote again by $H_\al$. Notice that in the above
equation $A= A(0)$ and $B=B(0)$.

We now  define  the trial function
\begin{multline}\label{sepp}
\Psi_\eps = \pe  \ua -\as \mA^{-1} (\vs \ua)\Ea \pe - d \as
\mA^{-1} \P\cdot\Da \pe - \al \mA^{-1} \Da\cdot\Da \pe +\\
 +\al\mA^{-1} (\vs \ua) \Ea \mA^{-1} (\vs \ua)\Ea \pe +2 \al \mA^{-1}
\P\Da \mA^{-1} (\vs \ua) \Ea \pe\;,
\end{multline}
with $\mA = P_f^2 + H_f$.

Comparing with the  minimizing sequence for $\so$ in
(\ref{tf2})--(\ref{tf}) we have replaced in (\ref{sepp}) the mere
electron function $f_n$ by $\pe$ and have added an extra
one-photon component $-d \as  \mA^{-1} \P\cdot\Da \pe$, which will
be responsible for lowering the energy, whereas the other one- and
two-photon parts will help to recover $\so$.

For short, we denote the 1- and 2- photons terms in $\Psi_\eps$ by
$\psi_1$ and  $\psi_2$ respectively. Obviously, the terms
$(\psi_1, P_f \cdot p \psi_1)$ and $(\psi_2, P_f \cdot p\psi_2)$
vanish, which can be immediately seen by integrating over the
field variables, having in mind (\ref{convention}) and the fact
that $\mA$ commutes with the reflection $k\to -k$.

By means of Schwarz' inequality and (\ref{pein}) we infer
\begin{multline}
\left| \Big([2\as p\cdot\Da + \as \vs\cdot \Ea]\as
\mA^{-1}p\cdot\Da\pe, \psi_2\Big)
\right| +\\
+ |(\psi_2,p_x^2 \psi_2)| \leq \|\Psi_\eps\|^2 \Ow(\al^{5/2}).
\end{multline}

Taking into account the negativity of $V$ and the estimates in the
proof of the upper bound in Section 2 we arrive at
\begin{multline}\label{seppi}
(\Psi_\eps, H_\al \Psi_\eps) \leq (\pe,[p^2 + V]\pe) - d \al (\pe,
p\cdot D
\mA^{-1}p\cdot\Da \pe)+ \\
+ \al d^2\Big[ (\pe, p\cdot D\mA^{-1} p\cdot\Da \pe) + (\pe,
p\cdot D\mA^{-1} p^2
\mA^{-1}p\cdot\Da \pe)\Big] + \\
+ [\so + \Ow\big(\al^{5/2}\ln(1/\al)\big)] \;\pa \Psi_\eps\pa^2.
\end{multline}
Using the Fourier transform we are able to evaluate explicitly
\begin{multline}
(\pe,p\cdot D\mA^{-1}p\cdot \Da \pe) = \sum_{\lambda =1,2} \int
|\hat \pe(l)|^2 \frac{[G^\lambda(p) \cdot l]^2}{|p|^2 + |p|} dp
dl= \\= \pa p \pe\pa^2 \pi^{-1} \int_0^{\la} \int_{-1}^1
\frac{\chi(|p|) x^2}{1 + |p|} dx d|p| = \frac{2}{3\pi} \ln(1+\la)
\pa p \pe\pa^2
\end{multline}
and analogously
\begin{multline}
(\pe,p\cdot D\mA^{-1}p^2\mA^{-1}p\cdot\Da \pe) = \frac{2}{3\pi}
\ln(1+\la)\pa p^2 \pe\pa^2 \\ \leq C_1 \frac{2}{3\pi} \ln(1+\la)
\pa p \pe\pa^2.
\end{multline}
Minimizing the corresponding terms in (\ref{seppi}) with respect
to $d$, leads to the requirement $d= \frac 1{2(C_1 +1)} $.

Finally it remains to choose an appropriate $\eps$ to guarantee
that
\begin{equation}\label{guar}
(\pe,[p^2 + V]\pe)-\al \frac{\ln(1+\la)}{6\pi(C_1+1)} \, \pa p
\pe\pa^2 < -\al \nu \pa p \psi \pa^2,
\end{equation}
for some $\nu(\eps) > 0$. By IMS localization formula (see e.g.
\cite[Theorem 3.2]{CF})
\begin{multline}
(\pe,[p^2 + V]\pe)= (\psi,[p^2 + V]\psi) -  (\psi v,[p^2 + V]\psi
v) \\+ (\psi, [|\nabla v|^2 + |\nabla u|^2]\psi ).
\end{multline}
The first term on the r.h.s. vanishes by assumption, the second
one is positive, and the third one is bounded by
\[
(\psi, [|\nabla v|^2 + |\nabla u|^2]\psi ) \leq C\, (\eps
\al)^2\int_{2 (\eps \al)^{-1} \geq |x| \geq (\eps \al)^{-1}} \frac
1{|x|^2} dx \leq C\, \,\al \,\eps\;,
\]
the constant depending on $\max \{|v'(t)|+ |u'(t)|\big| t \in
[1,2]\}$. Since
\begin{equation}
\pa p \pe\pa^2 \geq \pa p \psi\pa^2 - C\, \eps \al,
\end{equation}
we obtain (\ref{guar}) for  $\eps $ small enough. Consequently
\begin{equation}
(\Psi_\eps, H_\al\Psi_\eps)/(\Psi_\eps,\Psi_\eps)  \leq  -
\delta(\eps) \al^2 +\so+ \Ow\big(\al^{5/2}\ln(1/\al)\big)),
\end{equation}
which implies our claim.

\begin{appendix}
\section{Auxiliary operators}

For convenience we introduce the operators
\begin{eqnarray}\label{Da}
|D| &=& \sum_{\lambda =1,2} \int \frac{\chi(|k|)}{2\pi |k|^{1/2}}
a_\lambda (k) dk\;,\\
|E| &=& \sum_{\lambda =1,2} \int \frac{\chi(|k|)|k|^{1/2}}{2\pi }
a_\lambda (k) dk\;, \\ \label{Xa}
  |X| &=& \sum_{\lambda =1,2} \int \frac{\chi(|k|)}{2\pi\,|k|^{1/2}[|k| +
\al^3]^{1/2} } a_\lambda (k) dk\;.
\end{eqnarray}
 It is easily proved, using the commutation relations between the
 annihilation and creation operators, that
 \begin{equation}\label{xx*}
|X|\,|X|^\ast=|X|^\ast\,|X|+2\pi^{-1}\,\bigl(\la+3\al^{3}\,\ln(1/\al)-\al^{3}\,\ln(\la+\al^{3})
\bigr)\;.
 \end{equation}
\vskip6pt Moreover, analogously to \cite[Lemma A.4]{GLL} we obtain
the following.
\begin{lem}\label{auxop} For (\ref{Da})-(\ref{Xa}) we have
\begin{eqnarray}
|D|^*|D| &\leq& \frac 2\pi \la H_f\;;\\
|E|^*|E| &\leq & \frac {2\pi}3 \la H_f \;;\\  \label{rel3}
|X|^*|X| &\leq&  C\, \big [ |\ln(1/\al)| + |\ln(1 + \la)|\big]
H_f\;.
\end{eqnarray}
\end{lem}
\begin{rem} These newly defined operators now act on \textbf{real} functions.
Nevertheless to simplify the notation we shall often write
$|X|\psi$ instead of $|X|\,|\psi|$ for the $\C^2$-valued functions
we are considering.
\end{rem}
\begin{proof}
We only prove the inequality (\ref{rel3}). The proof for the other
terms  work similarly and is given in \cite[Lemma A.4]{GLL}.

Take an arbitrary $\Psi \in \Hh$ and fix the photons number $n$.
Then by means of Schwarz' inequality
\begin{multline}
(\psi_n,|X|^*|X| \psi_n) \leq 2 \Big( \int
\sqrt{\rho_{\psi_n}(k)|k|^{1/2} } \frac{\chi(|k|)}{|k|[|k| +
\al^3]^{1/2} } dk\Big)^2 \\ \leq  C\, \big [ |\ln(1/\al)| + |\ln(1
+ \la)|\big] \int \rho_{\psi_n}(k)|k| dk\;,
\end{multline}
since with the usual definition
\begin{equation}\label{A9}
\rho_{\psi_n}(k) = n\int |\psi_n(l,k,k_2,\dots,k_n)|^2 dl
dk_2\dots dk_n
\end{equation}
for the 1-photon density, we have
\begin{equation}\label{A10}
    \int_{\R^{3}}\rho_{\psi_n}(k)\,|
    k|\,dk=(\psi_{n},H_{f}\psi_{n})\;,
\end{equation}
while
\begin{equation}
   \label{log}
   \int \frac{\chi(|k_{n+1}|)^{2}}{|k_{n+1}|^{2}\,(|k_{n+1}|+\al^{3})}\,dk_{n+1}\sim \ln(1/\al)
\end{equation}
for $\al$ small enough.
\end{proof}
{}From now on, in order to lighten the notation,  $d^{n}k$ stands
for $dk_{1}\dots dk_{n}$.

\section{Evaluation of the contributing terms in (\ref{xian})}
Recall our notation
\begin{equation}
\P = p-P_f, \,\,\, F= 2\P\cdot D + \vs \cdot E\;.
\end{equation}
In the momentum representation of the electron space, $\P$ is
simply a multiplication operator and for short we use
\begin{equation}
\P \psi_n(l,k_1,\dots,k_n) = \Big(l-\sum_{i=1}^n k_i\Big) \psi_n
=: \P_n \psi_n\;,
\end{equation}
and similarly
\begin{equation}
H_f \psi_n(l,k_1,\dots,k_n) = \sum_{i=1}^n |k_i|\psi_n =: H_f^n
\psi_n\;.
\end{equation}
We shall also denote
\[
L_\al^n=|\P_n|^2 + H_f^n+\al^{3} \;.\]

For the sake of simplicity we will use in the following the
convention
\[|H|^2 := \sum_{\l = 1,2} |H^\l|^2, \,|G|^2 := \sum_{\l = 1,2} |G^\l|^2, \]
and additionally for all $a \in \R^3$
\[ |a\cdot G|^2 := \sum_{\l = 1,2} |a \cdot G^\l|^2. \]
These conventions are suggested by our definition of $H$ and $G$.
\vskip6pt

Before evaluating in  Lemma~\ref{lem:xian1} below the first term
in (\ref{xian1}), we need the following preliminary lemma.
\begin{lem}\label{lem:B1} For every $n\geq 0$,
\begin{multline}
\left|\,\| L_{\al}^{-1}\Fa\psi_{n}\|^{2} -\|\psi_{n}\|^{2}\;\|
\mA^{-1}\Ea\ora\|^{2}\,\right|\leq\\ \leq C\,\Big[\as
\|\psi_n\|^2+\al^{-1/2}\|\P\psi_n\|^2+\ln(1/\al)\,(\psi_n,
H_f\psi_n)\Big]\;.\label{lem:normeh}
\end{multline}
\end{lem}
\begin{proof}
  The l.h.s. of (\ref{lem:normeh}) is the sum of three terms~:
 \begin{multline}\label{dev}
\,\| L_\al^{-1}\Fa\psi_{n}\|^{2} =\|
L_\al^{-1}\vs\cdot\Ea\psi_{n}\|^{2}
+4\,\| L_\al^{-1}\P\cdot\Da\psi_{n}\|^{2}+\\
+4\,\Re (
L_\al^{-1}\vs\cdot\Ea\psi_{n},L_\al^{-1}\P\cdot\Da\psi_{n})\;.
\end{multline}
Each term is separately investigated in the three steps below.

\vskip6pt \noindent \textit{Step1.} The first term $\|
L_\al^{-1}\vs\cdot\Ea\psi_{n}\|^{2}$ is the one which contributes,
and we show that
\begin{multline*}
\left|\,\| L_{\al}^{-1}\vs\cdot\Ea\psi_{n}\|^{2}
-\|\psi_{n}\|^{2}\;\| \mA^{-1}\Ea\ora\|^{2}\,\right|\leq\\
\leq C\,\Big[\as \|\psi_n\|^2+\al^{-1/2}\|\P\psi_n\|^2+(\psi_n,
H_f\psi_n)\Big]\;.
\end{multline*}
This term  is decomposed into  a sum of two terms $I_{n}$ and
$II_{n}$, depending whether the same photon is created on both
sides or not.  Thanks to permutational symmetry and the
anti-commutation relations of the Pauli matrices, they are
respectively given by
 \begin{equation}\label{defIn}
 I_{n}=\int \frac{| H(k_{n+1})|^{2}\,|\psi_{n}(l,k_{1},\dots,k_{n})|^{2}}
 {\bigl(|\P_{n+1}|^{2}+H_{f}^{n+1}+\al^3\bigr)^{2}}\,dl dk_{1}\dots dk_{n+1}
     \end{equation}
and
\begin{multline}
    II_{n}=n \,\sum_{i,j=1}^{3}\int
    \frac{(\sigma_{j}\psi_{n}(l,k_{1},\dots,k_{n}),\sigma_{i}\psi_{n}(l,k_{2},\dots,k_{n+1}))}
 {\bigl(|\P_{n+1}|^{2}+H_{f}^{n+1}+ \al^3\bigr)^{2}}\times\\
 \times H_{j}(k_{n+1})\bar{H}_{i}(k_{1})\,dl dk_{1}\dots dk_{n+1}\;,
    \end{multline}
where the $\bar{\ }$ in the second line above refers to the
complex conjugate. We first evaluate $II_{n}$, for which it is
simply checked that
\begin{multline*}
II_{n}\leq\!\! \, \,C\,n\!\int \!\frac{|H(k_{1})|\,|H(k_{n+1})|}
{|k_{n+1}|\,|k_{1}|}\times \\ \times
|\psi_{n}(l,k_{1},\dots,k_{n})|\,
 |\psi_{n}(l,k_{2},\dots,k_{n+1})|\,dl\,d^{n+1}k\\
\leq \, \,
\!\!C\,\int\frac{\chi(|k|)}{|k|^2}\,dk\;(\psi_{n},H_{f}\psi_{n})\;,
\end{multline*}
thanks to \eqref{A9} and \eqref{A10}. We now examine
$I_{n}-\|\psi_{n}\|^{2}\;\| \mA^{-1}\Ea\ora\|^{2}$ and observe
that
\begin{equation*}
\|
\mA^{-1}\Ea\ora\|=\int_{\R^{3}}\frac{|H(k)|^{2}}{\big(|k|^{2}+|k|\big)^{2}}\,dk\;.
\end{equation*}
We first write
$L_{\al}^{n+1}=Q_{n+1}+|\P_{n}|^{2}+H_{f}^{n}+\al^3-2 \P_{n}\cdot
k_{n+1}$, with  $Q_{n+1}=|k_{n+1}|^{2}+|k_{n+1}|$. The following
quantity is then to be evaluated
\begin{multline*}
I_{n}-\|\psi_{n}\|^{2}\;\| \mA^{-1}\Ea\ora\|^{2}=\\
=\int |H(k_{n+1})|^{2}
\,|\psi_{n}(l,k_{1},\dots,k_{n})|^{2}\Big[\frac{1}{\big(L_{\al}^{n+1}\big)^2}-\frac{1}{Q_{n+1}{}^2}\Big]
 \,dl\,d^{n+1}k\;.
\end{multline*}
We now point out that
\begin{equation}\label{trick6}
\frac{1}{(Q+b)^2}=\frac{1}{Q^2}-\,\frac{2\,b}{Q\,(Q+b)^{2}}
-\,\frac{b^{2}}{Q^{2}\,(Q+b)^{2}}\;,
\end{equation}
apply this expression with $Q=Q_{n+1}+|\P_n|^2$ and
$b=H_{f}^{n}+\al^3-2 \P_{n}\cdot k_{n+1}$, and insert the
corresponding  expression into (\ref{defIn}). $I_n$ then appears
as a sum of three contributions
\[A_n=\int \frac{|H(k_{n+1})|^{2}}
 {\bigl(|\P_{n}|^{2}+Q_{n+1}\bigr)^{2}}|\psi_{n}(l,k_{1},\dots,k_{n})|^{2}\,dl d^{n+1}k\;,\]
\[B_n=2\,\int \frac{| H(k_{n+1})|^{2}\bigl(2\,\P_{n}\cdot k_{n+1}-H_{f}^{n}-\al^3\bigr)}
{(|\P_{n}|^{2}+Q_{n+1})\,\bigl(L_{\al}^{n+1}\bigr)^{2}}\,
|\psi_{n}(l,k_{1},\dots,k_{n})|^{2}\,dl d^{n+1}k\;, \] and
\[C_n=\int \frac{| H(k_{n+1})|^{2}\bigl(H_{f}^{n}+\al^3-2 \P_{n}\cdot k_{n+1} \bigr)^2}
 {(|\P_{n}|^{2}+Q_{n+1})^2\,\bigl(L_{\al}^{n+1}\bigr)^{2}}|\psi_{n}(l,k_{1},\dots,k_{n})|^{2}\,dl
d^{n+1}k\;.\]

First, applying again (\ref{trick6})  with $Q=Q_{n+1}$  and
$b=|\P_{n}|^{2}$,  it is a easily seen that
\[
\left|A_n- \|\psi_{n}\|^{2}\;\| \mA^{-1}\Ea\ora\|^{2}\right|\leq
C\,\int\frac{\chi(|k_{n+1}|)}{|k_{n+1}|^{2}}\,dk_{n+1}\,\|\P\psi_{n}\|^2\;,
\]
by using $\frac{|\P_n|^2}{|\P_{n}|^{2}+Q_{n+1}}\leq 1$. Concerning
$B_n$, we get on the one hand
\begin{multline*}
\int \frac{| H(k_{n+1})|^{2}\bigl(H_{f}^{n}+\al^3\bigr)}
 {(|\P_{n}|^{2}+Q_{n+1})\,\bigl(L_{\al}^{n+1}\bigr)^{2}}|\psi_{n}(l,k_{1},\dots,k_{n})|^{2}\,dl
d^{n+1}k\leq \\
\leq C\,\int\frac{\chi(|k_{n+1}|)}{|k_{n+1}|^{2}}\,dk_{n+1}\,\Big[
\big(\psi_n,H_f\psi_n\big)+\al^3\|\psi_n\|^2\Big]\;,
\end{multline*}
while, on the other hand, and with the help of  Schwarz'
inequality,
\begin{multline*}
\Big|\int \frac{| H(k_{n+1})|^{2}\,(\P_{n}\cdot k_{n+1})}
 {(|\P_{n}|^{2}+Q_{n+1})\,\bigl(L_{\al}^{n+1}\bigr)^{2}}|\psi_{n}(l,k_{1},\dots,k_{n})|^{2}\,dl
d^{n+1}k\Big|\leq\\
\leq C\,\int\frac{\chi(|k_{n+1}|)}{|k_{n+1}|}\,dk_{n+1}\,
\|\psi_n\|\,\|\P\psi_n\|\;. \end{multline*} For $C_n$, using
Young's inequality to deal with the cross term, we easily get
\begin{multline*}
|C_n|\leq
C\,\int\frac{\chi(|k_{n+1}|)}{|k_{n+1}|^{2}}\,dk_{n+1}\,\Bigl[
\big(\psi_n,H_f\psi_n\big)+\al^3\|\psi_n\|^2\Bigr]+\\
+C\,\int\frac{\chi(|k_{n+1}|)}{|k_{n+1}|}\,dk_{n+1}\,\|\P\psi_n\|^2\;,
\end{multline*}
since $\displaystyle{\frac{H_f^n+\al^3}{L_\al^{n+1}}\leq 1}$.

\vskip6pt \noindent \textit{Step 2.} We now show the following
bound on the second diagonal term~:
\begin{equation}
    \label{bornePD}
 (L_{\al}^{-1}\P\cdot\Da\psi_{n}, L_{\al}^{-1}\P\cdot\Da\psi_{n})
 \leq C\,\ln(1/\al)\,(\psi_{n},L\psi_{n}) \;.
\end{equation}
This quantity is again the sum of two terms $I_{n}+II_{n}$. We
first consider the ``diagonal" term $I_{n}$ for which the same
photon is created in both sides. It is worth observing that,
thanks to our choice of gauge for the potential vector $A$,
$G^\l(k)\cdot k=0$. Then, the  first term is bounded from above by
\begin{eqnarray*}
    I_{n}&\leq &\int \frac{|G(k_{n+1})|^{2}\,|\P_{n}|^{2} \,|\psi_{n}(l,k_{1},\dots,k_{n})|^{2}}
    {\Big(|\P_{n+1}|^{2}+H_{f}^{n+1}+\al^{3}\Big)^{2}}\,dl dk_{1}\dots dk_{n+1} \\
    &\leq&C\,\Big(\int \frac{|G(k_{n+1})|^{2}}{|k_{n+1}|\,(|k_{n+1}|+\al^{3})}\,dk_{n+1}\Big)
    \, \pa \P \psi_{n}\pa^{2}\\
    &\leq & C\,\ln(1/\al)\, \pa \P \psi_{n}\pa^{2}\;,
    \end{eqnarray*}
in virtue of \eqref{log}.

For the second term, we use $\frac {|\P|^2}{(|\P|^2 +
H_f+\al^{3})^2} \leq \frac 12 (H_f+\al^{3})^{-1}$ and  proceed as
follows
\begin{multline*}
 II_{n}\leq   n\sum_{\l=1,2} \int \frac{|G^\l(k_{n+1})|\,|\P_{n+1}|^{2} \,|G^\l(k_{1})|}
 {\Big(|\P_{n+1}|^{2}+H_{f}^{n+1}+\al^{3}\Big)^{2}}\times \\ \times
 |\psi_{n}(l,k_{1},\dots,k_{n})|\,|\psi_{n}(l,k_{2},\dots,k_{n+1})| \,dl\,
d^{n+1}k \\ \leq C\,(\psi_n,|X|^*|X|\psi_n) \leq C\,\ln(1/\al)\,
(\psi_{n}, H_{f}\psi_{n})\;,
    \end{multline*}
where the operator $|X|$ has been defined by (\ref{Xa}) in
Appendix A. (\ref{bornePD}) follows. \vskip6pt
\noindent\textit{Step 3.} Finally, we deal with the cross term in
(\ref{dev}) and show that
\begin{equation*}
|\Re( L_\al^{-1}\vs\cdot\Ea\psi_{n},L_\al^{-1}\P\cdot\Da\psi_{n})|
\leq C\, \big(\psi_{n},H_{f}\psi_{n}\big)\;.
  \end{equation*}
  Indeed, the term which corresponds to the case when one photon interacts with itself vanishes
  thanks to the fact that $G$ is real-valued while $H$ has purely imaginary
components. Observe now that, thanks to
\begin{equation}\label{trick1}
\frac {|\P|}{|\P|^2 + H_f+\al^3} \leq \frac
12\,(H_f+\al^3)^{-1/2}\leq \frac 12\,H_f^{-1/2}\;,
\end{equation}
\begin{equation*}
    \frac {|\P|}{(|\P|^2 + H_f+\al^{3})^2} \leq \frac 12
(H_f+\al^{3})^{-3/2}\leq \frac 12 H_f^{-3/2}\;,
\end{equation*}
and $\big(H_{f}^{n+1}\big)^{3/2}\geq
|k_{n+1}|^{5/4}\,|k_{1}|^{1/4}$. Then the remaining part gives
\begin{multline*}
| \Re(
L_\al^{-1}\vs\cdot\Ea\psi_{n},L_\al^{-1}\P\cdot\Da\psi_{n})| \leq
n \sum_{\l =1,2} \int
\frac{|H^\l(k_{n+1})|\,|\P_{n+1}|\,|G^\l(k_{1})|}
 {(L_\al^{n+1})^{2}}\, \times \\ \times|\psi_{n}(l,k_{1},\dots,k_{n})|\,|\psi_{n}(l,k_{2},\dots,k_{n+1})| \,dl\,
d^{n+1}k\\
\leq
C\int\frac{\chi(|k|)}{|k|^{5/2}}\,dk\,\big(\psi_n,H_{f}\psi_n\big)\,.
\end{multline*}
Lemma~\ref{lem:B1} follows collecting all above estimates.
\end{proof}
Let us now turn to the following.
\begin{lem}\textnormal{[\textbf{Evaluating the first term in} (\ref{xian1})]}\label{lem:xian1}
\begin{multline}
    \label{termxian1}
-\al \sum_{n\geq 0}\| L_\al^{-1/2}\Fa h_{n+1}\|^2=-\al \bigl(1-\|
\psi_{0}\|^{2}
\bigr)\,\lao E\mA^{-1}\Ea\ora +\\
+\al^{2}\lao E\mA^{-1} \Ea\ora \;\| \mA^{-1}\Ea\ora
\|^{2}+\Ow\big(\al^{5/2}\ln(1/\al)\big)\;.
    \end{multline}
\end{lem}
\begin{proof}
    As a direct consequence of (\ref{ordre1})
and (\ref{borneh}), we first get
\begin{multline}\label{inter}
-\al \sum_{n\geq 0}\| L_\al^{-1/2}\Fa h_{n+1}\|^2=\\=-\al\,\Big(
\sum_{n\geq 0} \| h_{n+1}\|^{2}\Big)\; \lao E\mA^{-1}\Ea \ora
+\Ow(\al^{3})\;.
\end{multline}
(Note that \eqref{ordre1} remains true with $L$ replaced with
$L_\al$.)  Next, we show that
\begin{equation}
    \sum_{n\geq 0}\| h_{n+1}\|^{2}=1-\| \psi_{0}\|^{2}-\al \| \mA^{-1}\Ea\ora \|^{2}
    +\Ow\big(\al^{3/2}\ln(1/\al)\big)\;.
    \label{eq:normeh}
\end{equation}
To this extent, using the definitions (\ref{defh1}) and
(\ref{defhn}) of $ h_{n+1}$, we get
\begin{eqnarray*}
    \lefteqn{\sum_{n\geq 0}\| \psi_{n+1}\|^{2}= 1-\| \psi_{0}\|^{2}=}  \\
     & = & \sum_{n\geq 0}\| h_{n+1}-\as\,L_\al^{-1}
\Fa\psi_{n}-\al\,L_\al^{-1}\Da\Da\psi_{n-1}\|^{2} \\
     & = &  \sum_{n\geq 0}\| h_{n+1}\|^{2}+\al\sum_{n\geq 0}
     \| L_\al^{-1}\Fa\psi_{n}\|^{2}-2\as \sum_{n\geq 0}\Re(h_{n+1},L_\al^{-1}
\Fa\psi_{n})- \\
     &  &-2\al \sum_{n\geq 0}\Re(h_{n+1},L_\al^{-1}
\Da\Da\psi_{n-1})+\Ow(\al^{3/2})\;,
\end{eqnarray*}
where $\Ow(\al^{3/2})$ comes both from the term
$\al^{2}\sum_{n\geq 0}
     \| L_\al^{-1}\Da\Da\psi_{n-1}\|^{2}$, and from the term
$\al^{3/2}\sum_{n\geq 0}\Re( L_\al^{-1}\Fa\psi_{n},
L_\al^{-1}\Da\Da\psi_{n-1})$, which is of the order of
$\al^{3/2}$, thanks to Schwarz' inequality and Lemma~\ref{lem:B1}
and the fact that
\begin{equation}
\| L_\al^{-1}\Da\Da\psi_{n-1}\|^{2}\leq C\big(\|\psi_{n-1}\|^2 +
\ln(1/\al)(\psi_{n-1},H_f\psi_{n-1})\big).
\end{equation}
Indeed, the diagonal part is obviously bounded by
\[\|\psi_{n-1}\|^2 \int\frac{|G(k_{n+1})|^2 \; |G(k_{n+1})|^2}{(|k_{n+1} |
+|k_{n+2}|)^2} dk_{n+1} dk_{n+2},\] whereas the off-diagonal part
is estimated by $(\psi_{n-1},|X|^*|X|\psi_{n-1})$.

With the help of Lemma \ref{lem:B1}  in Appendix~B, we have
\begin{displaymath}
 \al\sum_{n\geq 0}
     \| L_{\al}^{-1}\Fa\psi_{n}\|^{2}  = \al  \| \mA^{-1}\Ea\ora \|^{2}+\Ow(\al^{3/2})\;.
\end{displaymath}
Next, we prove that
\begin{equation}\label{bornehF}
\as \sum_{n\geq 0}| (h_{n+1},L_\al^{-1} \Fa\psi_{n})|\leq
C\,\al^{3/2}\ln(1/\al) \;.
\end{equation}
Let us indicate the main lines of the proof (\ref{bornehF}).
Thanks to the permutational symmetry of the photons variable, we
have
\begin{multline*}
| (h_{n+1},L_\al^{-1}
\Fa\psi_{n})|\leq \\
\leq \sqrt{n+1}\, \sum_{\l=1,2} \int\frac{
\big[2\,|G^\l(k_{n+1})\cdot \P_{n+1}|+|H^\l(k_{n+1})|\big]}{|\P_{n+1}|^{2}+H_{f}^{n+1}+\al^{3}}\times\\
\times |h_{n+1}(l,k_{1},\dots, k_{n+1})|
|\psi_{n}(l,k_{1},\dots,k_{n})|\,dldk_{1}\dots dk_{n+1}\;.
\end{multline*}
We begin with analyzing the term involving $H$ which appears to be
easier to deal with than the term involving $G$. This is due to
the two facts  that
\begin{equation}\label{trick8}
\frac{|H^\l(k_{n+1})|}{L_\al^{n+1}}\leq
C\,\frac{\chi(|k_{n+1}|)}{|k_{n+1}|^{1/2}}\;,
\end{equation}
whereas
\begin{equation}
\frac{|\P_{n+1}\cdot G^\l(k_{n+1})|}{L_\al^{n+1}}\leq
C\,\frac{\chi(|k_{n+1}|)}{|k_{n+1}|^{1/2}\,\big(|k_{n+1}|+\al^3\big)^{1/2}}
\label{trick7}
\end{equation}
in virtue of \eqref{trick1}.

On the one hand, using the fact that
$|\P_{n+1}|^{2}+H_{f}^{n+1}+\al^{3}\geq |k_{n+1}|$, the $H$-term
may be bounded by
\begin{multline}
    \sqrt{n+1}\sum_{\l=1,2}\int\frac{|h_{n+1}(l,k_{1},\dots, k_{n+1})|\,|H^\l(k_{n+1})| }
    {|\P_{n+1}|^{2}+H_{f}^{n+1}} \times\\\times
|\psi_{n}(l,k_{1},\dots,k_{n})|\,dl\,d^{n+1}k\\
\leq C\,\sqrt{n+1}\int |h_{n+1}(l,k_{1},\dots,
k_{n+1})|\,|k_{n+1}|^{1/2}\ \times
\\
\times|\psi_{n}(l,k_{1},\dots,k_{n})|\frac{ \chi(|k_{n+1}|)}{|k_{n+1}|}\,dld^{n+1}k\\
\leq C\, (h_{n+1}, H_{f}h_{n+1})^{1/2}\,\pa
\psi_{n}\pa\;,\label{bornehF1}
\end{multline}
thanks to Schwarz' inequality. On the other hand, for the
$G$-term, we shall make use of \eqref{trick1} to deduce the bound
\begin{multline}
\sqrt{n+1}\! \sum_{\l=1,2} \int\!\frac{|h_{n+1}(l,k_{1},\dots,
k_{n+1})|\, |G^\l(k_{n+1})\cdot
\P_{n+1}|}{|\P_{n+1}|^{2}+H_{f}^{n+1}+\al^{3}}
\,\times \\ \times|\psi_{n}(l,k_{1},\dots,k_{n})|\,dl\,d^{n+1}k\\
  \leq C\, \sqrt{n+1}\int\frac{|h_{n+1}(l,k_{1},\dots, k_{n+1})|\, |k_{n+1}|^{1/2}\,\chi(|k_{n+1}|)}
    {(|k_{n+1}|+\al^{3})^{1/2}\,|k_{n+1}|}\times\\
\times |\psi_{n}(l,k_{1},\dots,k_{n})|\,dl\,d^{n+1}k\\
\leq C\,(h_{n+1}, H_{f}h_{n+1})^{1/2}\,\Big(\int
\frac{ \chi(|k_{n+1}|)}{|k_{n+1}|^{2}\,(| k_{n+1}|+\al^{3})}\,dk_{n+1}\Big)^{1/2}\,\pa \psi_{n}\pa\\
\leq C\,\ln(1/\al)^{1/2}\;(h_{n+1}, H_{f}h_{n+1})^{1/2}\,\pa
\psi_{n}\pa\;,\label{bornehF2}
\end{multline}
thanks to \eqref{log}. Gathering together (\ref{bornehF1}) and
(\ref{bornehF2}), we deduce that
\begin{eqnarray*}
| (h_{n+1},L_\al^{-1}
\Fa\psi_{n})|&\leq&C\,\ln(1/\al)^{1/2}\;(h_{n+1}, H_{f}h_{n+1})^{1/2}\,\pa \psi_{n}\pa\\
&\leq& C\,\al\,\pa \psi_{n}\pa^{2}+C\ln(1/\al)\al^{-1}\,(h_{n+1},
H_{f}h_{n+1})\,;
\end{eqnarray*}
hence, (\ref{bornehF}) thanks to \eqref{borneh}.

Finally, we bound the last term in a similar way by
\begin{equation}\label{bornehDD}
 \al \sum_{n\geq 0}| (h_{n+1},L_{\al}^{-1}
\Da\Da\psi_{n-1})|\leq C\al^{2}\ln(1/\al)\;.
\end{equation}
Indeed, we recall that
\begin{multline*}
    \Da\cdot\Da\psi_{n-1}(l,k_{1},\dots,k_{n+1})
    =\frac{2}{\sqrt{n(n+1)}}\\
  \sum_{\l,\mu=1,2}  \sum_{i=1}^n\sum_{j=i+1}^{n+1}G^\l(k_{i})\cdot
    G^\mu(k_{j})\,\psi_{n-1}(l,k_{1},\dots,\check{k}_{i},\dots,
    \check{k}_{j},\dots, k_{n+1})\;.
\end{multline*}
Thus, thanks to permutational symmetry and since
$|\P_{n+1}|^{2}+H_{f}^{n+1}+\al^{3}\geq
2\,(|k_{n}|+\al^{3}/2)^{1/2}\,(|k_{n+1}|+\al^{3}/2)^{1/2}$, we may
bound this term as follows
\begin{eqnarray*}
\lefteqn{| (h_{n+1},L_{\al}^{-1}\Da\Da\psi_{n-1})|\leq}\\
&\leq& \sum_{\l,\mu =1,2} \sqrt{n(n+1)}\int
\frac{|G^\l(k_{n})|\,|G^\mu(k_{n+1})|}
{(|k_{n}|+\al^{3}/2)^{1/2}\,(|k_{n+1}|+\al^{3}/2)^{1/2}}\times\\
\, &&\times |\psi_{n-1}(l,k_{1},\dots,
k_{n-1})|\,|h_{n+1}(l,k_{1},\dots, k_{n+1})|\,dldk_{1}\dots
dk_{n+1}\\
&\leq & C\,(|X|\,|h_{n+1}|, |X|^\ast\,|\psi_{n-1}|)\\
&\leq & C\Big[
\al^{-1}\ln(1/\al)\,\big(h_{n+1},H_{f}h_{n+1}\big)\,+
\al\,\pa \psi_{n-1}\pa^{2}+\\
&&\mbox{\hskip4truecm}  +\al\ln(1/\al)\,
\big(\psi_{n-1},H_{f}\psi_{n-1}\big)\Big]\,,
\end{eqnarray*}
where the operator $|X|$ has been defined by (\ref{Xa}) in
Appendix A and where the last inequality follows from Schwarz'
inequality, (\ref{xx*}) and (\ref{rel3}). Hence (\ref{bornehDD})
thanks to (\ref{borneh}). \vskip6pt Hence (\ref{eq:normeh}).
Finally  (\ref{termxian1}) follows by inserting (\ref{eq:normeh})
into (\ref{inter}).
\end{proof}
We now prove the following
\begin{lem}\textnormal{[\textbf{Evaluating the second term in }\eqref{xian1}]}\label{lem:B2}
For every $n\geq 0$,
\begin{multline}
 \Bigl\vert\,
\pa L_{\alpha}^{-1/2}\Fa L_{\alpha}^{-1}\Fa\psi_{n}\pa^{2}
-\pa\psi_{n}\pa^{2}\;
\lao E \mA^{-1}E\mA^{-1}\Ea \mA^{-1}\Ea\ora\Bigr. -\\
\Bigl. -4\,\pa\psi_{n}\pa^{2}\;\lao E \mA^{-1}\P_{f}\cdot
D\mA^{-1}\P_{f}\cdot\Da
\mA^{-1}\Ea\ora\, \Big\vert\leq\\
\leq C\,\Big[\as \pa\psi_n\pa^2+\al^{-1/2}\pa
\P\,\psi_n\pa^2+\ln(1/\al)\big(\psi_n,L\psi_n\big)\Big]\;.
\end{multline}
\end{lem}
\begin{proof}

Thanks to the permutational symmetry, we have
\begin{multline}\label{asterm}
\pa L_{\alpha}^{-1/2}\Fa
L_{\alpha}^{-1}\Fa\psi_{n}\pa^{2}=\sum_{\l,\mu =1,2}\sum_{i=1}^{n+1}\sum_{j=i+1}^{n+2}\sum_{\gamma,\,\gamma',\,\nu,\nu'=1}^3
 \int\\ \Big(\frac{\big(\bar H^\l_\gamma(k_{n+2})\sigma_\gamma +2\P_{n+2}\cdot
G^\l(k_{n+2})\big)\big(\bar
H^\mu_{\gamma'}(k_{n+1})\sigma_{\gamma'}+2\P_{n+1}\cdot
G^\mu(k_{n+1})\big)}{L_\al^{n+2} \big(L_\al^{n+1}\big)^2} + \\ + \frac{\big(\bar H^\l_\gamma(k_{n+1})\sigma_\gamma +2\P_{n+2}\cdot
G^\l(k_{n+1})\big)\big(\bar
H^\mu_{\gamma'}(k_{n+2})\sigma_{\gamma'}+2\bar \P_{n+1}\cdot
G^\mu(k_{n+2})\big)}{L_\al^{n+2}\, L_\al^{n+1}\,\bar L_\al^{n+1}} \\
\psi_n(l,k_1,\dots,k_n), \,\big(H^\l_\nu(k_{i})\sigma_\nu
+\P_{n+1}\cdot
G^\l(k_{i})\big)\,\big(H^\mu_{\nu'}(k_{j})\sigma_{\nu'}+2\P_{n+2}\cdot
G^\mu(k_{j})\big)\\
\psi_n(l,k_1,\dots,\check{k_i},\dots,\check{k_j},\dots,
k_{n+2})\Big)\,dl\,d^{n+2}k\;,
\end{multline}
where $\bar \P_{n+1} = l - \sum_{i=1,\ne
n+1}^{n+2} k_i$ and $\bar L_\al^{n+1}= {\bar \P_{n+1}}^2 +  \sum_{i=1,\ne
n+1}^{n+2}|k_i| +\al^3$.
To avoid confusion corresponding to our notation we restrict our attention to the
first term in (\ref{asterm}). The proof of the second part works analogously.
The first quantity in  (\ref{asterm}) is decomposed in a sum of three terms $I_{n}$,
$II_{n}$ and $III_{n}$, which correspond respectively to the cases
$i=n+1$ and $j=n+2$, $i\not=n+1$ and $j=n+2$ and
$i,\,j\not\in\{n+1,n+2\}$. The terms will be respectively examined
in the three steps below. \vskip6pt \noindent\textit{Step 1.} We
first consider the diagonal term $I_n$. We use the fact that $H$
is complex valued while G is real valued to cancel all terms which
involve an odd number of $H$'s terms. In virtue of the
anti-commutation properties of the Pauli matrices,  we may write
\begin{eqnarray}
\lefteqn{I_n=\int\frac{|H(k_{n+2})|^{2}\,|H(k_{n+1})|^{2}}{L_\al^{n+2}\,
\big(L_\al^{n+1}\big)^2}\,|\psi_n(l,k_1,\dots,k_n)|^{2}\,dl\,d^{n+2}k+}\label{In1}\\
&+&4\int\frac{|H(k_{n+1})|^{2}\,|\P_{n+2}\cdot
G(k_{n+2})|^{2}}{L_\al^{n+2}\,
\big(L_\al^{n+1}\big)^2}\,|\psi_n(l,k_1,\dots,k_n)|^{2}\,dl\,d^{n+2}k+\label{In2}\\
&+&16 \int\frac{|\P_{n+1}\cdot G(k_{n+1})|^{2}\,|\P_{n+2}\cdot
G(k_{n+2})|^{2}}{L_\al^{n+2}\,
\big(L_\al^{n+1}\big)^2}\,|\psi_n(l,k_1,\dots,k_n)|^{2}\,dl\,d^{n+2}k+\nonumber\\
&+&4 \int\frac{|\P_{n+1}\cdot
G(k_{n+1})|^{2}\,|H(k_{n+2})|^{2}}{L_\al^{n+2}\,
\big(L_\al^{n+1}\big)^2}\,|\psi_n(l,k_1,\dots,k_n)|^{2}\,dl\,d^{n+2}k+\nonumber\\
&+&4\sum_{\l,\mu=1,2} \int\frac{\P_{n+1}\cdot
G^\l(k_{n+1})\,\P_{n+2}\cdot G^\mu(k_{n+2})\,H^\mu(k_{n+2})\cdot
H^\l(k_{n+1})}{L_\al^{n+2}\, \big(L_\al^{n+1}\big)^2}\times
\,\nonumber\\ && \qquad \times
|\psi_n(l,k_1,\dots,k_n)|^{2}\,dl\,d^{n+2}k.\nonumber
\end{eqnarray}
The first two terms will be the contributing ones and we leave
them temporarily apart. The three others are bounded as follows:
\begin{multline*}
\int\frac{|G(k_{n+1})|^{2}\,|\P_{n+2}|^{2}|G(k_{n+2})|^{2}}{L_\al^{n+2}\,
\big(L_\al^{n+1}\big)^2}\,|\P_{n}|^{2}\,
|\psi_n(l,k_1,\dots,k_n)|^{2}\,dl\,d^{n+2}k
\leq\\
\leq C\,
\big(\int\frac{\chi(|k|)}{|k|^{2}}\,dk\big)^2\;\|\P\psi_{n}\|^{2}\;,
\end{multline*}
by using that $\P_{n+1}\cdot G^\l(k_{n+1})=\P_{n}\cdot
G^\l(k_{n+1})$, similarly
\begin{multline*}
\int\frac{|\P_{n}|^{2}\,|G(k_{n+1})|^{2}\,|H(k_{n+2})|^{2}}{L_\al^{n+2}\,
\big(L_\al^{n+1}\big)^2}\,|\psi_n(l,k_1,\dots,k_n)|^{2}\,dl\,d^{n+2}k
\leq\\
\leq C\, \int\chi(|k_{n+2}|)\,dk_{n+2}\,
\int\frac{\chi(|k_{n+1}|)}{|k_{n+1}|^{2}\,\big(|k_{n+1}|+\al^{3}\big)}\,dk_{n+1}\,\|\P\psi_{n}\|^{2}\\
\leq C\, \ln(1/\al)\,\|\P\psi_{n}\|^{2}\;,
\end{multline*}
thanks to \eqref{log}, and
\begin{multline*}
\sum_{\l,\mu=1,2}
\int\!\frac{|H^\l(k_{n+2})|\,|H^\mu(k_{n+1})|\,|\P_{n}|\,|G^\mu(k_{n+1})|\,|\P_{n+1}|\,|G^\l(k_{n+2})|}{L_\al^{n+2}\,
\big(L_\al^{n+1}\big)^2}\, \times \\ \times
|\psi_n(l,k_1,\dots,k_n)|^{2}dld^{n+2}k
\\
\leq C\,\int\frac{\chi(|k_{n+2}|)}{|k_{n+2}|}\,dk_{n+2}\,
\int\frac{\chi(|k_{n+1}|)}{|k_{n+1}|^{3/2}}\,dk_{n+1}\,\|\psi_{n}\|\,\|\P\psi_{n}\|
\\
\leq C\,\big[\as\,
\|\psi_{n}\|^{2}+\al^{-1/2}\,\|\P\psi_{n}\|^{2}\big]\;,
\end{multline*}
with the help of \eqref{trick1}. We now turn to \eqref{In1} and
check that
\begin{multline}\label{evalIn1}
\Big|\int\frac{|H(k_{n+2})|^{2}\,|H(k_{n+1})|^{2}}{L_\al^{n+2}\,
\big(L_\al^{n+1}\big)^2}\,|\psi_n(l,k_1,\dots,k_n)|^{2}\,dl\,d^{n+2}k-\\
-\|\psi_n\|^2\int\!\!\frac{|H(k_{n+2})|^{2}|H(k_{n+1})|^{2}}
{Q_{n+2}\,
\big(Q_{n+1}\big)^2}\,dk_{n+1}dk_{n+2}\Big|\\
\leq C\,
\Big[\as\,\|\psi_n\|^2+\al^{-1/2}\,\|\P\psi_n\|^2+\ln(1/\al)\big(\psi_n,H_f\psi_n\big)\Big]
\;,\end{multline} with
$Q_{n+2}=|k_{n+2}+k_{n+1}|^2+|k_{n+2}|+|k_{n+1}|$ and
$Q_{n+1}=|k_{n+1}|^2+|k_{n+1}|$. Ob\-serve that
\begin{multline*}
\lao
E\mA^{-1}E\mA^{-1}\Ea\mA^{-1}\Ea\ora = \int\frac{|H(k_{n+2})|^{2}\,|H(k_{n+1})|^{2}} {Q_{n+2}\,
\big(Q_{n+1}\big)^2}\,dk_{n+1}dk_{n+2}+ \\ + \int\frac{|H(k_{n+2})|^{2}\,|H(k_{n+1})|^{2}} {Q_{n+2}\,
Q_{n+1}\big(|k_{n+2}|^2+|k_{n+2}|\big)}\,dk_{n+1}dk_{n+2}\;.
\end{multline*}
We first apply \eqref{trick6} to $\big(L_\al^{n+1}\big)^2$ with
$Q=Q_{n+1}+|\P_n|^2$ and $b=-2k_{n+1}\cdot\P_n+H_f^n+\al^3$. By
simple arguments which are very similar to those used  in the
course of the proof of Lemma~\ref{lem:B1} above (that we skip to
reduce the length of the calculations), we check that
\begin{multline*}
\Big|\int\frac{|H(k_{n+2})|^{2}\,|H(k_{n+1})|^{2}}{L_\al^{n+2}\,
\big(L_\al^{n+1}\big)^2}\,|\psi_n(l,k_1,\dots,k_n)|^{2}\,dl\,d^{n+2}k-\\
-\int\frac{|H(k_{n+2})|^{2}\,|H(k_{n+1})|^{2}}{L_\al^{n+2}\,
\big(Q_{n+1}+|\P_n|^2\big)^2}\,|\psi_n(l,k_1,\dots,k_n)|^{2}\,dl\,d^{n+2}k\Big|\leq \\
\leq C\,\Big[\as\,
\|\psi_{n}\|^{2}+\al^{-1/2}\,\|\P\psi_{n}\|^{2}+\big(\psi_{n},H_{f}\psi_{n}\big)\Big]\;.
\end{multline*} Next, we apply
\begin{equation}\label{trick9}
\frac{1}{Q+b}=\frac{1}{Q}-\frac{b}{Q\,(Q+b)}\;
    \end{equation}
to $L_{\al}^{n+2}$ with $Q=Q_{n+2}$ and
$b=-2(k_{n+2}+k_{n+1})\cdot\P_n+|\P_{n}|^{2}+H_f^n+\al^3$ and
obtain that
\begin{multline*}
\Big|\int\frac{|H(k_{n+2})|^{2}\,|H(k_{n+1})|^{2}}{L_\al^{n+2}\,
\big(Q_{n+1}+|\P_n|^2\big)^2}\,|\psi_n(l,k_1,\dots,k_n)|^{2}\,dl\,d^{n+2}k-\\
-\int\frac{|H(k_{n+2})|^{2}\,|H(k_{n+1})|^{2}}{Q_{n+2}\,
\big(Q_{n+1}+|\P_n|^2\big)^2}\,|\psi_n(l,k_1,\dots,k_n)|^{2}\,dl\,d^{n+2}k\Big|\leq \\
\leq C\,\Big[\as\,
\|\psi_{n}\|^{2}+\al^{-1/2}\,\|\P\psi_{n}\|^{2}+\big(\psi_{n},H_{f}\psi_{n}\big)\Big]\;.
\end{multline*} Finally, applying again \eqref{trick6} with
$Q=Q_{n+1}$ and $b=|\P_{n}|^{2}$, we get
\begin{multline*}
\Big|\int\frac{|H(k_{n+2})|^{2}\,|H(k_{n+1})|^{2}}{Q_{n+2}\,
\big(Q_{n+1}+|\P_n|^2\big)^2}\,|\psi_n(l,k_1,\dots,k_n)|^{2}\,dl\,d^{n+2}k-\\
-\int\frac{|H(k_{n+2})|^{2}\,|H(k_{n+1})|^{2}}{Q_{n+2}\,
\big(Q_{n+1}\big)^2}\,|\psi_n(l,k_1,\dots,k_n)|^{2}\,dl\,d^{n+2}k\Big|\leq \\
\leq C\,\Big[\as\,
\|\psi_{n}\|^{2}+\al^{-1/2}\,\|\P\psi_{n}\|^{2}+\big(\psi_{n},H_{f}\psi_{n}\big)\Big]\;.
\end{multline*}
The proof of \eqref{evalIn1} is then over and we now regard the
term in \eqref{In2} and show that
\begin{multline}\label{evalIn2}
\Big|\int\frac{|H(k_{n+1})|^{2}\,|\P_{n+2}\cdot
G(k_{n+2})|^{2}}{L_\al^{n+2}\,
\big(L_\al^{n+1}\big)^2}\,|\psi_n(l,k_1,\dots,k_n)|^{2}\,dl\,d^{n+2}k-\\
-\|\psi_n\|^2\int\!\!\frac{\big|(k_{n+2}+k_{n+1})\cdot
G(k_{n+2})\big|^{2}|H(k_{n+1})|^{2}} {Q_{n+2}\,
\big(Q_{n+1}\big)^2}\,dk_{n+1}dk_{n+2}\Big|\\
\leq C\,
\Big[\as\,\|\psi_n\|^2+\al^{-1/2}\,\|\P\psi_n\|^2+\big(\psi_n,H_f\psi_n\big)\Big]
\;,\end{multline} where
\begin{multline*}
\lao E\mA^{-1}\P_{f}\cdot
D\mA^{-1}\P_{f}\cdot\Da\mA^{-1}\Ea\ora=\\=\int\!\!\frac{\big|(k_{n+2}+k_{n+1})\cdot
G(k_{n+2})\big|^{2}|H(k_{n+1})|^{2}} {Q_{n+2}\,
\big(Q_{n+1}\big)^2}\,dk_{n+1}dk_{n+2}+\\
+ \int\!\!\frac{\big|(k_{n+2}+k_{n+1})\cdot
G(k_{n+2})\big|^{2}|H(k_{n+1})|^{2}} {Q_{n+2}\,Q_{n+1}
\big(|k_{n+2}|^2 + |k_{n+2}|\big)}\,dk_{n+1}dk_{n+2}
\end{multline*}
The proof is exactly the same as for \eqref{evalIn1}, therefore we
only sketch the main lines. Applying \eqref{trick6} to
$\big(L_\al^{n+1}\big)^2$ with $Q=|\P_n|^2+Q_{n+1}$ and
$b=-2\P_n\cdot k_{n+1}+H_f^n+\al^3$, we first arrive at
\begin{multline*}
\Big|\int\frac{|H(k_{n+1})|^{2}\,|\P_{n+2}\cdot
G(k_{n+2})|^{2}}{L_\al^{n+2}\,
\big(L_\al^{n+1}\big)^2}\,|\psi_n(l,k_1,\dots,k_n)|^{2}\,dl\,d^{n+2}k-\\
-\int\frac{|H(k_{n+1})|^{2}\,|\P_{n+2}\cdot
G(k_{n+2})|^{2}}{L_\al^{n+2}\,
\big(Q_{n+1}+|\P_n|^2\big)^2}\,|\psi_n(l,k_1,\dots,k_n)|^{2}\,dl\,d^{n+2}k\Big|\leq\\
\leq
C\,\big[\as\,\|\psi_n\|^2+\al^{-1/2}\,\|\P\psi_n\|^2+\big(\psi_n,H_f\psi_n\big)\Big]
\,.\end{multline*} Next, again from \eqref{trick9}, with
$Q=Q_{n+1}$ and $b=|\P_n|^2$, we obtain
\begin{multline*}
\Big|\int\frac{|H(k_{n+1})|^{2}\,|\P_{n+2}\cdot
G(k_{n+2})|^{2}}{L_\al^{n+2}\,
\big(Q_{n+1}+|\P_n|^2\big)^2}\,|\psi_n(l,k_1,\dots,k_n)|^{2}\,dl\,d^{n+2}k-\\
-\int\frac{|H(k_{n+1})|^{2}\,|\P_{n+2}\cdot
G(k_{n+2})|^{2}}{L_\al^{n+2}\,
\big(Q_{n+1}\big)^2}\,|\psi_n(l,k_1,\dots,k_n)|^{2}\,dl\,d^{n+2}k\Big|\leq\\
\leq C\,\|\P\psi_n\|^2 \,,\end{multline*} and we  use
\eqref{trick9} with $Q=Q_{n+2}$ and $b=-2\P_n\cdot
(k_{n+1}+k_{n+2})+|\P_n|^2+H_f^n+\al^3$ to get
\begin{multline*}
\Big|\int\frac{|H(k_{n+1})|^{2}\,|\P_{n+2}\cdot
G(k_{n+2})|^{2}}{L_\al^{n+2}\,
\big(Q_{n+1}\big)^2}\,|\psi_n(l,k_1,\dots,k_n)|^{2}\,dl\,d^{n+2}k-\\
-\int\frac{|H(k_{n+1})|^{2}\,|\P_{n+2}\cdot
G(k_{n+2})|^{2}}{Q_{n+2}\,
\big(Q_{n+1}\big)^2}\,|\psi_n(l,k_1,\dots,k_n)|^{2}\,dl\,d^{n+2}k\Big|\leq \\
\leq
C\,\big[\as\,\|\psi_n\|^2+\al^{-1/2}\,\|\P\psi_n\|^2+\big(\psi_n,H_f\psi_n\big)\Big]
\,.\end{multline*} Finally, since
$\P_{n+2}=\P_n-(k_{n+1}+k_{n+2})$ and $G^\l(k_{n+2})\cdot
k_{n+2}=0$, we obtain
\begin{multline*}
\int\frac{|H(k_{n+1})|^{2}\,|\P_{n+2}\cdot
G(k_{n+2})|^{2}}{Q_{n+2}\,
\big(Q_{n+1}\big)^2}\,|\psi_n(l,k_1,\dots,k_n)|^{2}\,dl\,d^{n+2}k=\\
=\|\psi_n\|^{2}\,\int\frac{|H(k_{n+1})|^{2}\,|(k_{n+1}+k_{n+2})\cdot
G(k_{n+2})|^{2}}{Q_{n+2}\,
\big(Q_{n+1}\big)^2}\,dk_{n+1}dk_{n+2}+\\
+2\sum_{\l=1,2}\int\frac{|H(k_{n+1})|^{2}\big(k_{n+1}\cdot
G^\l(k_{n+2})\big)\big(\P_n\cdot G^\l(k_{n+2})\big)}{Q_{n+2}\,
\big(Q_{n+1}\big)^2} \times \\ \times|\psi_n(l,k_1,\dots,k_n)|^{2}dl\,d^{n+2}k+\\
+\int\frac{|H(k_{n+1})|^{2}\,|\P_n\cdot G(k_{n+2})|^{2}}{Q_{n+2}\,
\big(Q_{n+1}\big)^2}\,|\psi_n(l,k_1,\dots,k_n)|^{2}\,dl\,d^{n+2}k
\,.
\end{multline*}
The second term in the  r.h.s. vanishes when integrated first with
respect to $k_{n+1}$ since $H$ and $Q_{n+1}$ are radially
symmetric functions, whereas the second term is easily bounded by
\[
C\,\int\frac{\chi(|k_{n+2}|)}{|k_{n+2}|^2}\,dk_{n+2}\;\int\frac{\chi(|k_{n+1}|)}{|k_{n+1}|}\,dk_{n+1}\;
\|\P\psi_n\|^2\;.
\]

This concludes the proof of \eqref{evalIn2}. \vskip6pt
\noindent\textit{Step 2.} We now regard the term  $II_{n}$ which,
thanks to permutational symmetry,  can be bounded by
\begin{multline*}
|II_{n}|\leq C\,n\!\!\sum_{\l,\mu=1,2}
\int\!\!\frac{\big(|H^\mu(k_{n+2})|+2|\P_{n+2}\cdot
G^\mu(k_{n+2})|\big)^{2}}{L_\al^{n+2}\, \big(L_\al^{n+1}\big)^2}
\times \\ \times\,\big(|H^\l(k_{n+1})|+2|\P_{n+1}\cdot
G^\l(k_{n+1})|\big) \big(|H^\l(k_{1})|+2|\P_{n+1}\cdot
G^\l(k_{1})|\big)\times
\\ \times
|\psi_n(l,k_1,\dots,k_n)|\,|\psi_n(l,k_2,\dots,
k_{n+1})|\,dl\,d^{n+2}k\;.
\end{multline*}
We are going to show that
\begin{equation*}
|II_n|\leq C\,\ln(1/\al)(\psi_n,H_f\psi_n)\;.
\end{equation*}
 First observe that it
is enough to study the case of \\ $|H(k_{n+2})|^2+4|\P_{n+2}\cdot
G(k_{n+2})|^{2}$. Since
\[\frac{|H(k_{n+2})|^{2}}{L_\al^{n+2}\big(L_\al^{n+1}\big)^2}\leq
C\,\frac{\chi(|k_{n+2}|)}{\big(L_\al^{n+1}\big)^2}, \] whereas,
using $\P_{n+2}\cdot G^\l (k_{n+2})=\P_{n+1}\cdot G^\l (k_{n+2})$,
\[
\frac{|\P_{n+2}\cdot G
(k_{n+2})|^{2}}{L_\al^{n+2}\big(L_\al^{n+1}\big)^2}\leq
C\,\frac{\chi(|k_{n+2}|)}{|k_{n+2}|^2\,L_\al^{n+1}}\;,
\]
in virtue of \eqref{trick1}, it is easily seen that the $|H|^2$
contribution is the most delicate to handle since it involves a
higher power of $|k_1|+|k_{n+1}|$ at the denominator. We thus
concentrate on this term. Moreover, comparing \eqref{trick8} and
\eqref{trick7} it is easily seen that the ``worse" term may be
bounded as follows
\begin{multline*}
n\,\sum_{\l=1,2} \int\frac{|\P_{n+1}\cdot
G^\l(k_{n+1})|\,|\P_{n+1}\cdot
G^\l(k_{1})|}{\big(L_\al^{n+1}\big)^2}\times\\
 \times |\psi_n(l,k_1,\dots,k_n)|\,|\psi_n(l,k_2,\dots,
k_{n+1})\,dl\,d^{n+1}k\leq \\
\leq
C\,n\,\sum_{\l=1,2}\int\frac{|G^\l(k_{n+1})|\,|G^\l(k_{1})|}{|k_{n+1}|^{1/2}\,
\big(|k_{n+1}|+\al^3\big)^{1/2}|k_{1}|^{1/2}\,\big(|k_{1}|+\al^3\big)^{1/2}}\,\times
\\
\times |\psi_n(l,k_1,\dots,k_n)|\,|\psi_n(l,k_2,\dots,
k_{n+1})|\,dl\,d^{n+1}k\leq\\
\leq C\,\ln(1/\al)(\psi_n,H_f\psi_n)\;,
\end{multline*}
thanks to Schwarz' inequality and \eqref{log}.

 \vskip6pt
\noindent\textit{Step 3.} We finally consider the full
off-diagonal term that we first roughly bound by
\begin{multline*}
|III_{n}|\leq C\,n(n-1)\!\!\sum_{\l,\mu=1,2} \\
\int\!\!\frac{\big(|H^\l(k_{n+2})|+|\P_{n+2}\cdot
G^\l(k_{n+2})|\big)\,\big(|H^\mu(k_{n+1})|+|\P_{n+1}\cdot
G^\mu(k_{n+1})|\big)}{L_\al^{n+2}\, \big(L_\al^{n+1}\big)^2}\times\\
\times \big(|H^\l(k_{1})|+|\P_{n+2}\cdot G^\l(k_{1})|\big) \,
\big(|H^\mu(k_{2})|+|\P_{n+1}\cdot
G^\mu(k_{2})|\big)\times\\\times
|\psi_n(l,k_1,\dots,k_n)|\,|\psi_n(l,k_3,\dots,
k_{n+2})|\,dl\,d^{n+2}k\;.
\end{multline*}

The term only involving the $H$'s is bounded by
\begin{multline*}
|III_{n}|\leq C\,n(n-1)\!\!\sum_{\l,\mu=1,2}
\int\!\!\frac{|H^\l(k_{n+2})|\,|H^\mu(k_{n+1})|\,|H^\l(k_{1})||H^\mu(k_{2})|}{H_f^{n+1}
|k_2||k_{n+1}|}\times \\ \times
|\psi_n(l,k_1,\dots,k_n)|\,|\psi_n(l,k_3,\dots,
k_{n+2})|\,dl\,d^{n+2}k\;\\ \leq
C\||E|H_f^{-1/2}|D||\psi_n|\|^2\leq C(\psi_n, H_f\psi_n),
\end{multline*}
and the corresponding term with the $G$'s reads
\begin{multline*}
|III_{n}|\leq C\,n(n-1)\!\!\sum_{\l,\mu=1,2}  \int\!\!\frac{|
G^\l(k_{n+2})|\,| G^\mu(k_{n+1})|\,| G^\l(k_{1})|\,|
G^\mu(k_{2})|\big)}{L_\al^{n+2}\, \big(L_\al^{n+1}\big)^2}\times
\\ \times |\P_{n+1}|^2 \, |\P_{n+1}|^2
\,|\psi_n(l,k_1,\dots,k_n)|\,|\psi_n(l,k_3,\dots,
k_{n+2})|\,dl\,d^{n+2}k\; \\ \leq C\,n(n-1)\!\!\sum_{\l,\mu=1,2}
\int\!\!\frac{| G^\l(k_{n+2})|\,| G^\mu(k_{n+1})|\,|
G^\l(k_{1})|\,| G^\mu(k_{2})|\big)}{L_\al^{n+1}}\times \\\times
|\psi_n(l,k_1,\dots,k_n)|\,|\psi_n(l,k_3,\dots,
k_{n+2})|\,dl\,d^{n+2}k \\ \leq C \||D|H_f^{-1/2}|D||\psi_n|\|^2
\leq C(\psi_n, H_f\psi_n).
\end{multline*}
The mixed terms then are estimated by means of Schwarz'
inequality.
\end{proof}
Finally, we recover the last contributing term by proving the
following.
\begin{lem}\textnormal{[\textbf{Evaluating the term in }(\ref{xian3})]}\label{lem:B3}
For every $n\geq 0$,
\begin{multline}
\Big|\,\Re (L_{\al}^{-1}\Fa
L_{\al}^{-1}\Fa\psi_{n},\Da\Da\psi_{n})-
\|\psi_{n}\|^{2}\;\lao E \mA^{-1}E\mA^{-1}\Da \Da\ora\,\Big|\leq\\
\leq C\,\Big[\al^{-1/2}\ln(1/\al)\,(\psi_n,L\psi_n)+\as \,\pa
\psi_n\pa^2\Big]\;.
\end{multline}
\end{lem}
\begin{proof}
\vskip6pt\noindent\textit{Step 1.} We first observe  that, by
Schwarz' inequality,
\begin{eqnarray*}
    \lefteqn{| (L_{\al}^{-1}\Fa L_{\al}^{-1}\P\cdot
    \Da\psi_{n},\Da\Da\psi_{n})|\leq}\\
    &\leq& C\, \pa  L_{\al}^{-1}\P\cdot \Da\psi_{n}\pa\,\pa F L_{\al}^{-1}\Da\Da\psi_{n}\pa\\
    &\leq & C\,\al^{-1/2} \pa  L_{\al}^{-1}\P\cdot
    \Da\psi_{n}\pa^{2}+C\,
    \as\, \pa F L_{\al}^{-1}\Da\Da\psi_{n}\pa^2\\
    &\leq& C\,\big[\al^{-1/2}\ln(1/\al)\,(\psi_n,L\psi_n)+\as\,\pa
    \psi_{n}\pa^{2}+\as \,(\psi_n,H_{f}\psi_n)\Big]\;,
\end{eqnarray*}
thanks to (\ref{bornePD}) and since the other $\Ll^2$ norm is
easily checked to be bounded due to the fact that
\begin{equation*}
\Fa F\leq C\, (H_f + |\P|^2\,H_f)
\end{equation*}
in virtue of \cite[Lemma A.4]{GLL}.

\vskip6pt\noindent\textit{Step 2.} We now look at the term
\begin{multline*}
   \Re(L_{\al}^{-1}\P\cdot
    \Da L_{\al}^{-1}\vs\cdot \Ea \psi_{n},\Da\Da\psi_{n})= 2\sum_{\l,\mu=1,2}
\sum_{\gamma=1}^{3}\times\\ \times
 \Re \int\frac{\P_{n+2}\cdot G^\l(k_{n+2})\,
  \bar H^{\mu}_{\gamma}(k_{n+1})\,\sum_{i=1}^{n+1}\sum_{j=i+1}^{n+2}G^\l(k_{i})\cdot
  G^\mu(k_{j})}{\big[|\P_{n+1}|^2 +
H_f^{n+1}+\al^3\big]\,\big[|\P_{n+2}|^2 + H_f^{n+2}+\al^3\big]}\times\\
  \times \big(\vs_{\gamma}\psi_{n}(l,k_{1},\dots,k_{n}),
  \psi_{n}(l,k_{1},\dots,\check{k}_{i},\dots,\check{k}_{j},\dots,
  k_{n+2})\big)\,dld^{n+2}k\;.
\end{multline*}
The  diagonal term, when $i=n+1$ and $j=n+2$, vanishes since $H$
is purely imaginary while $G$ is real. We then have three
off-diagonal terms to deal with, $I_{n}$, $II_{n}$ and $III_{n}$,
which correspond respectively to the cases $j=n+2$, $j=n+1$ and
$j\not\in\{n+1,n+2\}$.

First, using (\ref{trick1}) and $\frac{|
H^\l(k_{n+1})|}{|\P_{n+1}|^2 + H_f^{n+1}+\al^3}\leq |
G^\l(k_{n+1})|$,
\begin{eqnarray*}
| I_{n}| &\leq& n\sum_{\l=1,2} \int\frac{| \P_{n+2}| \, |
G(k_{n+2})|^2\,
  | H^{\l}(k_{n+1})| \, | G^\l(k_{1})|}{\big[|\P_{n+2}|^2 +
  H_f^{n+2}+\al^3\big]\,\big[|\P_{n+1}|^2 +
H_f^{n+1}+\al^3\big]}\times\\
  &&\times  | \psi_{n}(l,k_{1},\dots,k_{n})|\,
  | \psi_{n}(l,k_{2},\dots, k_{n+1})| \,dld^{n+2}k\\
  &\leq & C\, \int\frac{| G(k_{n+2})|^{2}}{|
  k_{n+2}|^{1/2}}\,dk_{n+2}\ \pa | D|
  \,|\psi_{n}| \,\pa^{2}\leq C\, (\psi_{n},H_{f}\psi_{n})\;,
 \end{eqnarray*}
thanks to Lemma \ref{auxop} and \eqref{trick1}. Secondly, thanks
again to (\ref{trick1}) and Lemma~\ref{auxop}, we have
\begin{eqnarray*}
| II_{n}| &\leq& n \sum_{\l,\mu=1,2} \int\frac{ | G^\l(k_{n+2})|\,
  | H^\mu(k_{n+1})| | G^\mu(k_{n+1})|\, | G^\l(k_{1})|}{\big[
  H_f^{n+2}+\al^3\big]^{1/2}\,\big[
H_f^{n+1}+\al^3\big]}\times\\
  &&\times  | \psi_{n}(l,k_{1},\dots,k_{n})|\,
  | \psi_{n}(l,k_{2},\dots, \check{k}_{n+1},k_{n+2})| \,dl\,d^{n+2}k\\
  &\leq & C\,\sum_{\l=1,2} \int\frac{| G^\l(k_{n+1})|\,| H^\l(k_{n+1})|}{|
  k_{n+1}|^{3/2} }\,dk_{n+1}\ \pa | D|
  \,|\psi_{n}|\pa^{2} \\
  &\leq& C\,\big(\psi_{n},H_{f}\psi_{n}\big)\;.
 \end{eqnarray*}
Finally, the full off-diagonal term reads
\begin{eqnarray*}
| III_{n}| &\leq& n\,(n-1)\sum_{\l,\mu=1,2} \int\frac{ |
G^\l(k_{n+2})|\,
  | H^\mu(k_{n+1})|\, | G^\l(k_{1})|\, | G^\mu(k_{2})|}{\big[
  H_f^{n+2}+\al^3\big]^{1/2}\,\big[
H_f^{n+1}+\al^3\big]}\times\\
  &&\times  | \psi_{n}(l,k_{1},\dots,k_{n})|\,
  | \psi_{n}(l,k_{3},\dots,k_{n+2})| \,dldk_{1}\dots dk_{n+2}\\
  &\leq & C\, \big|(|X|\,H_{f}^{-1/2}
  \,| D| \ \psi_{n} ,\, | D| H_{f}^{-1/2}
  \,|E|\ \psi_{n})\big|\\
  &\leq& C\,\ln(1/\al)^{1/2} (\psi_{n},H_{f}\psi_{n})\;.
\end{eqnarray*}

\vskip6pt\noindent\textit{Step 3.} To conclude the proof of the
lemma, we are thus lead to prove that
\begin{multline*}\Big|\,\Re
(L_{\al}^{-1}\vs\cdot\Ea
L_{\al}^{-1}\vs\cdot\Ea\psi_{n},\Da\Da\psi_{n})-
\|\psi_{n}\|^{2}\;\lao E \mA^{-1}E\mA^{-1}\Da \Da\ora\,\Big|\leq\\
\leq C\, \as \|\psi_{n}\|^{2}+C\,\al^{-1/2}(\psi_n,L\,\psi_n)\;.
\end{multline*}
On the one hand, using the explicit formulations of the operators
$E$, $D$ and their adjoints, we recall that
\begin{multline*}
\lao E \mA^{-1}E\mA^{-1}\Da
\Da\ora=\\
=2\,\sum_{\l,\mu=1,2} \,\int_{\R^{3}\times\R^{3}}\frac{ \bar
H^\l(k_{1})\cdot \bar H^\mu(k_{2})\; G^\l(k_{1})\cdot
G^\mu(k_{2})}{\big[|k_{1}|^{2}+|k_{1}|\big]\,\big[|k_{1}+k_{2}|^{2}+|k_{1}|+|k_{2}|\big]}
\,dk_{1}dk_{2}\;.
\end{multline*}
On the other hand
\begin{multline*}
\Re(L_{\al}^{-1}\vs\cdot\Ea
L_{\al}^{-1}\vs\cdot\Ea\psi_{n},\Da\Da\psi_{n})=\\
=2\sum_{\l,\mu=1,2} \Re \sum_{\gamma,\gamma'=1}^{3}\int\frac{\bar
H^\l_{\gamma}(k_{n+2})\,
  \bar{H}^\mu_{\gamma'}(k_{n+1})\,\sum_{i=1}^{n+1}\sum_{j=i+1}^{n+2}G^\l(k_{i})\cdot
  G^\mu(k_{j})}{\big[|\P_{n+2}|^2 + H_f^{n+2}+\al^3\big]\,\big[|\P_{n+1}|^2 +
H_f^{n+1}+\al^3\big]}\times\\
   \times (\vs_{\gamma}\psi_{n}(l,k_{1},\dots,k_{n}),
  \vs_{\gamma'}\psi_{n}(l,k_{1},\dots,\check{k}_{i},\dots,\check{k}_{j},\dots,
  k_{n+2}))\,dl\,d^{n+2}k\;.
\end{multline*}
This term may again be decomposed as a sum of three terms
according to the same convention as above. Nevertheless it is
easily checked that only the first term, which corresponds to
$i=n+1$ and $j=n+2$, contributes, while the other ones may be
bounded from above by exactly the same method as before. Following
the scheme of proof of Lemma~\ref{lem:B1} and Lemma~\ref{lem:B3},
we introduce further simplifying notation~:
\[R_{n+2}=L_{\al}^{n+2}-\Q_{n+2}=-2\P_{n}\cdot
(k_{n+1}+k_{n+2})+|\P_{n}|^{2}+H_{f}^{n}+\al^{3}\;,
\]
and
\[
R_{n+1}=L_{\al}^{n+1}-\Q_{n+1}=-2\P_{n}\cdot
k_{n+1}+|\P_{n}|^{2}+H_{f}^{n}+\al^{3}\;.
\]
The following difference is then to be evaluated
\begin{multline}\label{diff}
\sum_{\l,\mu=1,2}
\int\Big[\frac{1}{L_{\al}^{n+2}\,L_{\al}^{n+1}}-\frac{1}{\Q_{n+2}\,\Q_{n+1}}\Big]\,
 \bar H^\mu(k_{n+2})\cdot \bar H^\l(k_{n+1})\,\times\\
 \times  G^\l(k_{n+1})\cdot G^\mu(k_{n+2})\, |\psi_{n}(l,k_{1},\dots,
  k_{n})|^{2}\,dldk_{1}\dots dk_{n+2}\;.
\end{multline}
It is straightforward to check that
\begin{subequations}
    \label{trick4}
\begin{eqnarray}
 \lefteqn{\frac{1}{L_{\al}^{n+2}\,L_{\al}^{n+1}}-\frac{1}{\Q_{n+2}\,\Q_{n+1}}=}\nonumber\\
    &=&2\frac{\P_{n}\cdot (k_{n+2}+k_{n+1})}{L_{\al}^{n+1}\,\Q_{n+2}\,\Q_{n+1}}
    +2\frac{\P_{n}\cdot
    k_{n+1}}{L_{\al}^{n+2}\,\Q_{n+2}\,\Q_{n+1}}-\label{trick41}\\
    &&-\,\Big[\frac{L_{\al}^{n}}{L_{\al}^{n+1}\,\Q_{n+2}\,\Q_{n+1}}+
   \frac{L_{\al}^{n}}{L_{\al}^{n+2}\,\Q_{n+2}\,\Q_{n+1}}\Big]+\label{trick42}\\
   &&+\frac{R_{n+1}\,R_{n+2}}{L_{\al}^{n+1}\,L_{\al}^{n+2}\,\Q_{n+2}\,\Q_{n+1}}\;.\label{trick43}
\end{eqnarray}
\end{subequations}
We now insert this expression into (\ref{diff}) and simply bound
$|G^\l(k_{n+1})| \times$ $\times|H^\l(k_{n+1})|$ by
$C\,\chi(|k_{n+1}|)$ and similarly for
$|G^\mu(k_{n+2})|\,|H^\mu(k_{n+2})|$.  It is then very easy to
bound the two terms in (\ref{trick41}) by $C\, \pa\psi_n\pa \,
\pa\P\psi_n\pa$ and the terms in (\ref{trick42}) by $C\,
(\psi_n,L\psi_n)+C\, \al^3 \pa\psi_n\pa^2$. Concerning
(\ref{trick43}), the term involving
$\frac{|\P_n|^2\,|k_{n+1}|\,|k_{n+1}+k_{n+2}|}
{L_{\al}^{n+1}\,L_{\al}^{n+2}\,\Q_{n+2}\,\Q_{n+1}}$ is also easily
bounded by $\pa\P\psi_n\pa^2$ while all the terms involving
$H_f^n+\al^3$ admit simple bounds by $C\, \pa\psi_n\pa \,
\pa\P\psi_n\pa$ or $C\, (\psi_n,L\psi_n)+C\, \al^3
\pa\psi_n\pa^2$. To deal with the remaining terms
\begin{equation}\label{trash}
\frac{2|\P_n|\,|k_{n+1}|\,|\P_n|^2}{L_{\al}^{n+1}\,L_{\al}^{n+2}\,\Q_{n+2}\,\Q_{n+1}},\;
\frac{2|\P_n|\,|\P_n|^2\,|k_{n+1}+k_{n+2}|}{L_{\al}^{n+1}\,L_{\al}^{n+2}\,\Q_{n+2}\,\Q_{n+1}}\;,
\frac{|\P_n|^4}{L_{\al}^{n+1}\,L_{\al}^{n+2}\,\Q_{n+2}\,\Q_{n+1}}\;,
\end{equation}
we observe that, from \eqref{trick9},
\begin{equation}\label{trick5}
\frac{1}{L_{\al}^{n+2}}=\frac{1}{L_{\al}^{n}+\Q_{n+2}}-\,\frac{-2\P_{n}\cdot(k_{n+1}+k_{n+2})
+|k_{n+1}+k_{n+2}|^{2}}{L_{\al}^{n+2}\,\big(L_{\al}^{n}+\Q_{n+2})}\;.
\end{equation}
Since  $L_{\al}^{n}=|\P_{n}|^{2}+H_{f}^{n}+\al^{3}$, inserting
(\ref{trick5}) in (\ref{trash}) and using the two bounds
\[ \frac{|\P_{n}|^{2}}{L_{\al}^{n}+\Q_{n+2}}\leq 1\; \textrm{ and }\;
\frac{|\P_{n}|}{L_{\al}^{n}+\Q_{n+2}}\leq \frac{1}{2\,\big
(H_{f}^{n}+\al^{3}+\Q_{n+2}\big)^{1/2}} \;,\] it is a tedious but
easy exercise to bound the contribution of all the terms in
(\ref{trash}) by $\pa \P\psi_{n}\pa^{2}$, except for one term
which comes from the last term in (\ref{trash}) and  which is
precisely bounded by
\[
\frac{|\P_n|^5\,|k_{n+1}+k_{n+2}|}{L_{\al}^{n+1}\,L_{\al}^{n+2}(L_{\al}^{n}+\Q_{n+2})\,\Q_{n+2}\,\Q_{n+1}}\;.
\]
To handle this term, we plug in (\ref{trick5}) once more, and with
the  same two bounds as above, we again bound the contribution by
$\pa \P\psi_{n}\pa^{2}$.

\vskip6pt We now turn to the bound on the non-contributing terms.
Using first that $L_{\al}^{n+1}\,L_{\al}^{n+2}\geq |k_{n+1}|^{2}$,
we check that
\begin{multline*}
| II_{n}| \leq n\sum_{\l,\mu=1,2} \int\frac{| H^\l(k_{n+1})| |
G^\l(k_{n+1})|}{|k_{n+1}|^{2}}\,|H^\mu(k_{n+2})|\,|
G^\mu(k_{1})|\times\\\times | \psi_{n}(l,k_{2},\dots,
\check{k}_{n+1},k_{n+2})|\,| \psi_{n}(l,k_{1},\dots,k_{n})|\,
   \,dldk_{1}\dots dk_{n+2}\\
  \leq  C\,\sum_{\l=1,2} \int\frac{| G^\l(k_{n+1})|\,| H^\l(k_{n+1})|}{ |k_{n+1}|^{2} }\,dk_{n+1}
  \big| (| D|\,|\psi_{n}| ,\,| E| \,|\psi_{n}|)\big|\\
  \leq C\, (\psi_{n},H_{f}\psi_{n})\;,
 \end{multline*}
while, with $L_{\al}^{n+1}\,L_{\al}^{n+2}\geq
|k_{n+2}|\,(\sum_{i=3,\,\not
=n+1}^{n+2}|k_{i}|)^{1/2}\,(\sum_{i=2}^{n}|k_{i}|)^{1/2}$, we have
\begin{eqnarray*}
| III_{n}| &\leq& \displaystyle{n (n-1)\sum_{\l,\mu=1,2}
\int\frac{ | H^\l(k_{n+2})|\,
  |H^\mu(k_{n+1})|}{|k_{n+2}|\,(\sum_{i=3}^{n+2}|k_{i}|)^{1/2}}\,| \psi_{n}(l,k_{3},\dots,k_{n+2})\times}\\
 &&\times\frac{| G^\l(k_{1})|\, | G^\mu(k_{2})|}{(\sum_{i=2}^{n}|k_{i}|)^{1/2}}
   \,| \psi_{n}(l,k_{1},\dots,k_{n})|\,dldk_{1}\dots dk_{n+2}\\
  &\leq & C\,\big| (| D| H_{f}^{-1/2}|E|\,\psi_{n} ,|D|H_{f}^{-1/2}|D| \,\psi_{n})\big|\\
  &\leq& C\, (\psi_{n},H_{f}\psi_{n})\;.
 \end{eqnarray*}
\end{proof}

\section{Evaluation of the terms of higher order in (\ref{xian})}

First, we investigate the cross-terms in (\ref{xian}) which appear
with a factor $\al^{3/2}$.

\begin{lem}\textnormal{[\textbf{Bound on
}(\ref{xian4})]}\label{lem:C1}
\begin{multline}\label{equ1l1}
\big|\big(L_\al^{-1}F^*L_\al^{-1}F^*\psi_n,F^*h_{n+1}\big)\big|
\leq C\, \Big[\al \|\psi_n\|^2 + \al\big(\psi_n,H_f\psi_n\big)+\al
\|\P\psi_n\|^2\\ + \al^{-1}\big(h_{n+1},H_f h_{n+1}\big)\Big].
\end{multline}
\end{lem}
\begin{proof}
For shortness we restrict ourselves to the case $F=2\P\cdot D$,
which is the most delicate one. The other cases work similarly.

By permutational symmetry the first part of the l.h.s. of
(\ref{equ1l1}) is bounded from above by
\begin{multline}
\sqrt{n+1} \sum_{\lambda,\mu=1,2} \int
\frac{\big[G^\lambda(k_{n+2}) \cdot \P_{n+2}\big]^2
\big|G^\mu(k_{n+1}) \cdot \P_{n+1}\big|} { \big[|\P_{n+2}|^2 +
H_f^{n+2}+\al^3\big]\,\big[|\P_{n+1}|^2 + H_f^{n+1}+\al^3\big]}
\times \\ \times
|\psi_n(l,k_1,\dots,k_n)||h_{n+1}(l,k_1,\dots,k_{n+1})| dl
d^{n+2}k \\ \leq \sum_{\lambda=1,2}\int
\frac{|G^\lambda(k_{n+2})|^2}{|k_{n+2}|} dk_{n+2}
\;\big|\big(\P\psi_n,|D|
h_{n+1}\big)\big|\\
\leq C\, \|\P\psi_n\|\,\big(h_{n+1},H_fh_{n+1}\big)^{1/2}\;,
\end{multline}
since $G^\l(k_{n+1})\cdot\P_{n+1}=G^\l(k_{n+1})\cdot\P_{n}$ and
where we used (\ref{trick1}) and additionally $\frac {\P^2}{\P^2 +
H_f} \leq 1$.

The second, off-diagonal, part can be estimated by
\begin{eqnarray}
\Big|\big( |D|\psi_n,|D|H_f^{-1/2}|D|h_{n+1}\big)\Big|&\leq&
C\,\big(\psi_{n}, H_{f}\psi_{n}\big)^{1/2}\,\big(h_{n+1},
H_{f}h_{n+1}\big)^{1/2}\\
&\leq & C\,\Big[\al \big(\psi_{n},
H_{f}\psi_{n}\big)+\al^{-1}\big(h_{n+1}, H_{f}h_{n+1}\big)\Big]\,,
\nonumber\end{eqnarray} again with Schwarz' inequality and Lemma
\ref{auxop}.
\end{proof}

\begin{lem}\textnormal{[\textbf{Bound on }\eqref{xian5}]}\label{lem:C2}
\begin{multline}
\big|\big(L_\al^{-1}D^*D^*\psi_n,F^*h_{n+1}\big)\big| \leq C\,
\Big[\al \pa\psi_n\pa^2 +\as
\big(\psi_n,H_f\psi_n\big)+\\
+\al^{-1}\big(h_{n+1},H_f h_{n+1}\big)\Big]\;.
\end{multline}
\end{lem}
\begin{proof}
We restrict once again to $F=2\P\cdot D $. The absolute value of
the diagonal part is bounded by
\begin{multline}
\sqrt{n+1}\, \sum_{\lambda,\mu=1,2} \int
\frac{\big|G^\lambda(k_{n+2}) \cdot
\P_{n+2}\big|\,|G^\lambda(k_{n+2}) |\,|G^\mu(k_{n+1})|}
{\big[|\P_{n+2}|^2 + H_f^{n+2}+\al^3\big]} \times \\
\times |\psi_n(l,k_1,\dots,k_n)||h_{n+1}(l,k_1,\dots,k_{n+1})| dl
d^{n+2}k \\
\leq \sum_{\lambda=1,2}\int
\frac{|G^\lambda(k_{n+2})|^2}{|k_{n+2}|^{1/2}} dk_{n+2}
\Big|\big(\psi_n,|D| h_{n+1}\big)\Big|\\
\leq C\, \pa \psi_{n}\pa\,\big(h_{n+1}, H_{f}h_{n+1}\big)^{1/2}\;,
\end{multline}
with the help of (\ref{trick1}), whereas the off-diagonal term can
again be bounded by
\begin{equation}
\Big|\big( |D|\psi_n,|D|H_f^{-1/2}|D|h_{n+1}\big)\Big|\leq C\,
\big(\psi_{n}, H_{f}\psi_{n}\big)^{1/2}\big(h_{n+1},
H_{f}h_{n+1}\big)^{1/2}\,.
\end{equation}
\end{proof}

For the term appearing with $\al^2$ in (\ref{xian8}) we derive
\begin{lem}\textnormal{[\textbf{Bound on }(\ref{xian8})]}\label{lem:C3}
\begin{multline}
\big|\big(L_\al^{-1}\Fa L_\al^{-1}\Da\Da\psi_{n-1},\Fa
h_{n+1}\big) \big|\leq C\,\Big[\al \|\psi_{n-1}\|^2 +
\big(\psi_{n-1},H_f\psi_{n-1}\big)+\\
+\al^{-1}\ln(1/\al)\big(h_{n+1},H_f h_{n+1}\big)
+\bigl(h_{n+1},H_f h_{n+1}\bigr)\Big]\;.
\end{multline}
\end{lem}
\begin{proof}
Consider again $F = 2 \P\cdot D$. The main term reads
\begin{multline*}
(n+1) \sum_{\lambda,\mu,\nu=1,2}\int \frac{\big[G^\lambda(k_{n+2})
\cdot \P_{n+2}\big]^2 \big|G^\mu(k_{n+1}) \big|\big|G^\nu(k_{n})
\big|}{ \big[|\P_{n+1}|^2 +
H_f^{n+1}+\al^{3}\big]\,\big[|\P_{n+2}|^2 +
H_f^{n+2}+\al^{3}\big]}
\times\\
\times
|\psi_{n-1}(l,k_1,\dots,k_{n-1})||h_{n+1}(l,k_1,\dots,k_{n+1})| dl
d^{n+2}k\\
\leq C\,(|X|^*\psi_{n-1},|X|h_{n+1})\\
\leq C\,\ln(1/\al)^{1/2}\big(h_{n+1},H_f
h_{n+1}\big)^{1/2}\Big[\pa \psi_{n-1}\pa
+\\+\ln(1/\al)^{1/2}\big(\psi_{n-1},H_f
\psi_{n-1}\big)^{1/2}\Big],
\end{multline*}
whereas the totally off-diagonal term can be estimated by
\begin{multline*}
    \big|(|D|\psi_{n-1},
|D|H_f^{-1/2}|D|H_f^{-1/2}|D|h_{n+1})\big|\leq\\
\leq C\,\big(\psi_{n-1},H_f
\psi_{n-1}\big)^{1/2}\,\big(h_{n+1},H_f h_{n+1}\big)^{1/2}\,.
\end{multline*}
\end{proof}

In the following we consider the cross terms in (\ref{xian}) which
appear with a factor $\al^{5/2}$, for  which a rough estimate is
enough. Therefore we merely indicate the proofs.
\begin{lem}\textnormal{[\textbf{Bound on }(\ref{xian6})]}\label{lem:C4}
\begin{multline}
\big|\big(L_\al^{-1}F^*L_\al^{-1}F^*\psi_n,F^*L_\al^{-1}
D^*D^*\psi_{n-1}\big)\big| \leq C\, \Big[\as\,\|\psi_{n-1}\|^2 +
\as\,\pa \psi_n\pa ^{2}+\\ +\al^{-1/2}\,\big(\psi_n,H_f\psi_n\big)
+ \al^{-1/2}\,\big(\psi_{n-1},H_f \psi_{n-1}\big)\Big].
\end{multline}
\end{lem}
\begin{proof}
We restrict again to $F= 2\P\cdot D$ and regard only one diagonal
term, namely
\begin{multline}
(n+1)^{1/2} \sum_{\lambda,\mu,\nu=1,2} \int
\frac{\big[G^\lambda(k_{n+2}) \cdot \P_{n+2}\big]^2
\big|G^\mu(k_{n+1}) \cdot \P_{n+1}\big|}{ \big[|\P_{n+1}|^2 +
H_f^{n+1}+\al^{3}\big]^2\,\big[|\P_{n+2}|^2 +
H_f^{n+2}+\al^{3}\big]} \times \\ \times
  \big|G^\mu(k_{n+1})\big|
\big|G^\nu(k_{n})\big|
|\psi_n(l,k_1,\dots,k_n)||\psi_{n-1}(l,k_1,\dots,k_{n-1})| dl
d^{n+2}k \\ \leq \sum_{\lambda,\mu=1,2}\int
\frac{|G^\mu(k_{n+1})|^2}{|k_{n+1}|^{3/2}}{|G^\lambda(k_{n+2})|^2}
dk_{n+1}dk_{n+2} \big|\big(\psi_{n-1},|D| \psi_n\big)\big|\\
\leq C\,\pa \psi_{n-1}\pa\, (\psi_n,H_f\psi_n\big)^{1/2}.
\end{multline}
The remaining terms are estimated similarly.
\end{proof}
By similar methods the following concluding lemma concerning the
error term (\ref{xian7}) is obtained.
\begin{lem}\textnormal{[\textbf{Bound on }(\ref{xian7})]}\label{lem:C5}
\begin{multline}
\big| \big (L_\al^{-1}\Fa
L_\al^{-1}\Da\Da\psi_{n-1},\Da\Da\psi_{n})|
  \big| \leq C\,
\Big[\as\,\|\psi_{n-1}\|^2 + \as\,\pa \psi_n\pa ^{2}+\\
+\al^{-1/2}\,\big(\psi_n,H_f\psi_n\big) +
\al^{-1/2}\,\big(\psi_{n-1},H_f \psi_{n-1}\big)\Big].
\end{multline}
\end{lem}
Notice that in the last two lemmas simple Schwarz estimates would
suffice.
\end{appendix}

\vskip20pt \noindent\textbf{Acknowledgements.} C.H. has been
supported by a Marie Curie Fellowship of the European Community
programme \lq\lq Improving Human Research Potential and the
Socio-economic Knowledge Base\rq\rq\ under contract number
HPMFCT-2000-00660. This work has been partially financially
supported by the ACI project  of the French Minister for Education
and Re\-search.

\end{document}